\documentclass[12pt]{article}
\usepackage{amsmath, amsthm, amsfonts, bm}
\usepackage{amssymb}
\usepackage{mathtools}
\usepackage{float}

\usepackage{hyperref}
\hypersetup{colorlinks=true, citecolor=blue, linkcolor=blue, filecolor=blue}
\usepackage{graphicx,psfrag,epsf, float}
\usepackage{enumerate,titlesec, color}
\usepackage{natbib}
\usepackage{footnote}
\usepackage{url} 
\usepackage{soul}
\usepackage{tikz}
\usepackage{algorithm, algorithmicx, algpseudocode}

\usepackage[font=small,labelfont=bf]{caption}
\usepackage{subcaption}

\usepackage{algorithm, algorithmicx, algpseudocode}
\usepackage{booktabs}
\usepackage{lineno}

\usepackage{soul}

\usepackage{setspace}

\usepackage[textwidth=6.5in,bottom=1.8in,top=1in]{geometry}

\renewcommand\footnotemark{}

\usepackage{booktabs}

\usepackage{array}
\def\bftau{\mathbf \tau}
\usepackage{multirow}

\newcommand{\cm}[1]{\ignorespaces}

\def\bfa{\mathbf a}

\def\bfx{\mathbf x}

\def\bfA{\mathbf A}

\def\bfE{\mathbf E}

\def\bfalpha{\boldsymbol \alpha}
\def\bfbeta{\boldsymbol \beta}

\def\bfgamma{\boldsymbol \gamma}

\def\bfPhi{\boldsymbol \Phi}

\def\bfzero{\boldsymbol 0}

\def\bfTheta{{\ensuremath\boldsymbol{\Theta}}}

\begin{document}
	\bigskip
	\date{}

\title{Bayesian pathway analysis over brain network mediators for survival data}
\author{Xinyuan Tian$^{1}$, Fan Li$^{1,2}$, 
Li Shen$^{3}$, Denise Esserman$^{1,2}$ and     Yize Zhao$^{1,2,*}$ \\
\small$^{1}$Department of Biostatistics, Yale School of Public Health, New Haven, CT, US.\\
\small$^{2}$Yale Center for Analytical Sciences, New Haven, CT, US.\\
\small$^{3}$Department of Biostatistics, Epidemiology and Informatics, University of Pennsylvania, \\ \small Philadelphia, PA, US.}
\maketitle

\def\spacingset#1{\renewcommand{\baselinestretch}%
	{#1}\small\normalsize} \spacingset{1}

\begin{abstract}
Technological advancements in noninvasive imaging facilitate the construction of whole brain interconnected networks, known as brain connectivity. Existing approaches to analyze brain connectivity frequently disaggregate the entire network into a vector of unique edges or summary measures, leading to a substantial loss of information. 
Motivated by the need to explore the effect mechanism among genetic exposure, brain connectivity and time to disease onset, we propose an integrative Bayesian framework to model the effect pathway between each of these components while quantifying the mediating role of brain networks. To accommodate the biological architectures of brain connectivity constructed along  white matter fiber tracts, we develop a structural modeling framework which includes a symmetric matrix-variate accelerated failure time model and a symmetric matrix response regression to characterize the effect paths. We further impose within-graph sparsity and between-graph shrinkage to identify informative network configurations and eliminate the interference of noisy components. Extensive simulations confirm the superiority of our method compared with existing alternatives. By applying the proposed method to the landmark Alzheimer's Disease Neuroimaging Initiative  study, we obtain neurobiologically plausible insights that may inform future intervention strategies.
\end{abstract}

\noindent%
{\textbf{Keywords}: Accelerated failure time model; 
Brain connectivity; Imaging genetics; Markov chain Monte Carlo; Natural indirect effect; Shrinkage and regularization.}

\newpage
\spacingset{1.45}

\section{Introduction}
The field of neuroscience acknowledges that a nervous system is defined by interconnections of its neuronal units \citep{bassett2017network}. Through advances in noninvasive imaging techniques,  one could establish a cartography of these neuronal interconnections through large-scale networks, known as brain connectivity. Among different types of brain connectivity, structural connectivity is cutting-edge, defined by white matter fiber traits measured via diffusion magnetic resonance imaging (dMRI). Compared with regional neuroimaging measurements that characterize brain structure or function at separate locations, brain connectivity provides unique descriptions of neuronal dynamic patterns, and is expected to align with different 
types of brain activities \citep{jirsa2007handbook}.

Though a flurry of successes have been achieved to relate brain connectivity with human behavior \citep{shen2017using,greene2018task}, most of these studies were based on healthy individuals. It is highly likely that important disease metrics, such as time to disease onset, will  be impacted by brain alternations and their relationship may also be shaped by a potential exposure  factor. The methods proposed in this article are motivated by Alzheimer's disease (AD), a neurodegenerative disease that causes brain atrophy to destroy cognition. Besides being associated with a variety of changes in the brain, AD is also highly heritable with genetic risk factors including apolipoprotein E (APOE) genotyping strongly implicating disease progression \citep{bertram2008thirty}. To characterize the relationship between AD outcomes and genetic biomarkers, recent advances in neuroimaging genetics manage to substitute disease phenotypes with imaging endophenotypes to explore the associated genetic variants \citep{hashimoto2015imaging}. However, the pathological mechanism among genetic factors, neuroimaging traits and disease outcomes remains unclear.

To uncover the effect mechanism among the genetic exposure, brain structural connectivity and time to AD onset, we propose a unified Bayesian mediation framework with a time-to-event outcome and a network-variate mediator.  Mediation analysis has been developed to understand the mechanisms relating to how one phenomenon exerts its influence on another \citep{mediation2008}. By partitioning the overall effect on a dependent variable into a direct effect coming from an independent variable and an indirect effect passing through a mediator, one is able to dissect the effect pathways functioning among different components. The initial implementations of mediation analysis focus on a single mediator with a continuous outcome \cite{imai2013identification}, with extensions to accommodate multivariate or high-dimensional mediators \citep{song2020bayesian, derkach2019high, huang_and_pan_2016} and other types of outcome, including binary \citep{wang_2013} and time-to-event \citep{SurMed2017}; please also see \cite{zeng2021statistical} for a review on modeling development for mediation analysis.  Most of the studies applying these methods to neuroimaging data consider mediating traits in a scalar or vector form summarized by regional imaging measurements \citep{lindquist2012functional,zhao2019granger,chen2018high,zhao2021multimodal}. 
More recently, \citet{zhao2022bayesian} made the first attempt to handle brain functional connectivity as a unified mediator. However, they consider a continuous outcome and assume a stochastic block structure for the mediating networks, which is not applicable for our motivating example which involves  brain structural connectivity with  distinct network architectures and a time-to-event outcome subject to right censoring.

To properly model this complicated neurobiological mechanism, it is essential to characterize the unique topological form of structural connectivity with biological interpretability. Current analyses on structural connectivity primarily transfer the whole network into a vector of unique edges or summary measures \citep{simpson2011exponential,ballester2017whole,chen2022imaging}, resulting in a loss of information. \cite{wang2021learning} summarize the connectivity via a number of small clique subgraphs, which aligns with the biological architecture of white matter fiber traits. Taking this idea further, we propose that structural connectivity mediates the relationship through a set of clique signaling subgraphs that are influenced by the genotype and impact the outcomes. To accommodate the symmetric and hollow structure of connectivity, we propose a structural modeling framework which includes a symmetric matrix-variate accelerated failure time (AFT) model for the effect pathway from brain connectivity to the outcome, and a symmetric matrix response regression for the effect pathway from the genetic exposure to brain connectivity. Under a unified Bayesian paradigm, to identify informative brain networks along each effect pathway, we impose a within-graph sparsity to select significant network configurations, as well as a between-graph shrinkage to eliminate interference from noisy elements. To the best of our knowledge, each of the above modeling components itself has not yet been considered in previous brain imaging studies. 

Our contributions in this work are multi-fold. First, we provide the first attempt to develop a mediation analysis for a time-to-event outcome mediated by a network-variate mediator. We propose innovative modeling strategies to accommodate the biological configurations for brain structural connectivity, and characterize its mediating effect between the genetic exposure and time to disease onset. Second, we simultaneously dissect the network architecture functioning along each effect pathway within our structural modeling, and identify separate sets of informative subgraphs linked with the exposure and outcome. Third, we develop a unified Bayesian inference strategy with biologically plausible priors and allow for efficient posterior inference. The Bayesian approach allows us to coherently 
quantify the uncertainty for each effect element under the proposed structural models, and can be advantageous over the frequentist for complex modeling frameworks as in our setting.

The remainder of article is organized as follow. In Section \ref{sec: method}, we introduce a new network-variate mediation framework with symmetric and hollow structural constraints. In Section \ref{sec:bayesian}, we operate under a full Bayesian framework, discuss the prior specifications and develop the posterior inference procedures. We conduct simulation studies to evaluate the model performance in Section \ref{sec:simu}, and implement our method to study effect pathways among the genetic exposure, brain connectivity and AD survival in Section \ref{sec:ADNI} using the Alzheimer’s Disease Neuroimaging Initiative (ADNI) study. We conclude with a discussion in Section \ref{sec:dis}.

\section{Methods} \label{sec: method}

\subsection{Model formulation}
We start with a general model formulation. For subject $i~ (i=1,\dots,N)$, let $T_{i}$ denote the time to disease onset, and $C_{i}$ the right censoring time. The observed follow-up time can be represented as $\widetilde{T}_{i}=\min \{T_{i}, C_{i}\}$, and we define a censoring indicator $\delta_{i}=1$ if an event is observed and 0 if censored. Let $z_i$ denote a binary exposure, which in our application is the AD risk genotype, and $\bfx_i\in \mathbb{R}^{Q\times 1}$ represents a set of covariates to be adjusted for, including an intercept. We assume the brain structural connectivity can be summarized by a graph $\mathcal{G}_{i}=(\mathcal{R}, \mathcal{E}_{i})$ with a common set of nodes $\mathcal{R}$ and subject-specific edges  $\mathcal{E}_{i}$. Given the graphs are defined under the same brain atlas, the shared set of nodes is defined by a total of $R$ brain regions of interest (ROIs) within which we extract data from voxels. Subsequently, we  represent $\mathcal{G}_{i}$ by a symmetric connectivity matrix $\boldsymbol{A}_{i}\in\mathbb{R}^{R\times R}$ with its $(w,l)$th entry $a_{iwl}$ characterizing the white matter fiber tracts between ROIs $w$ and $l$, $0< w \neq l \leq R$. By construction, $\boldsymbol{A}_{i}$ is a hollow matrix with $a_{iwl}=0$, if $0< w = l\leq R$. 

Our goal is to characterize the effect pathways among the time-to-event outcome, exposure and brain structural connectivity. Simultaneously, we aim to identify brain subgraph configurations that function along each effect path where the neural system is involved. As stated in the previous section, it is important to maintain the topological architectures of brain connectivity. On the one hand, different from social networks, each connectivity matrix $\boldsymbol{A}_{i}$ is symmetric and with a hollow structure corresponding to the bidirectional and no self-connections of  structural connectivity. On the other hand, it is well accepted that brain connectivity operates through strongly interrelated subgraphs to offer effective neural support for a behavior \citep{bassett2017network}.  In light of those considerations, we propose the following structural models
\begin{equation}
\begin{aligned}\log(T_i)&=\bfx^T_i\bfbeta_x+\sum_{j=1}^{J}\big\langle\omega_j\bfbeta_j\bfbeta^T_j, \bfA_i\big\rangle_F+\beta_zz_i+\epsilon_i;\label{eq:aft}
\end{aligned}
\end{equation}
    \vspace{-0.6cm}
\begin{equation}
\begin{aligned}
\bfA_i&=\boldsymbol{G}_i-\mbox{Dg}[\boldsymbol{G}_i]+(\bfE_i-\bfE_i^T) \\
\boldsymbol{G}_i&=\mathcal{M}\times_3\bfx^T_i+\sum_{h=1}^{H}\eta_h\bfalpha_h\bfalpha^T_hz_i.
    \label{eq:matrix}
\end{aligned}
\end{equation}
Here, model \eqref{eq:aft} captures the effects on the survival outcome through an AFT model with $\bfbeta_x \in \mathbb{R}^{Q\times 1}$ and $\beta_z$ representing the effects of covariates and exposure on the logarithmic survival time, respectively. We assume $\epsilon_i\sim \mbox{N}(0, \sigma_0^2)$ in our application without loss of generality, as  other parametric distributions such as log-normal or extreme value distributions could also be adopted with minor modifications. For the mediating effect, we assume the structural connectivity impacts the outcome through a combination of $J$ network configurations, which we refer to as subgraphs. Here, $\langle\cdot,\cdot\rangle_{F}$ denotes the Frobenius inner product, and we characterize the effect on the outcome from subgraph $j$ through a matrix $\omega_j\bfbeta_j\bfbeta^T_j$ with coefficients $\bfbeta_j \in \mathbb{R}^{R\times 1}$ describing the within subgraph effect of each node, and $\omega_j$ capturing the impact between subgraphs. Given different brain network configurations are employed under distinct neuronal processes, 
it is anticipated each $\bfbeta_j$ is sparse to induce a specific clique subgraph contributing to the effect pathway on the outcome. From an analytical perspective, introducing sparsity will  reduce the burden of estimating parameters over a high-dimensional feature space. Technical details on sparsity will be discussed in Section \ref{sec:prior}. 

Joint with model \eqref{eq:aft}, model \eqref{eq:matrix}, which consists of two sub-models, captures the impact from the exposure to the connectivity mediator. With $A_i$ on the left-hand side of the equation, we essentially have a network response regression, where the network is constrained by a symmetric and hollow structure. To align with the zero diagonal elements of $A_i$, we first denote the matrix-variate effect from the covariates and exposure as $\boldsymbol{G}_i$. Within $\boldsymbol{G}_i$, $\mathcal{M}\in \mathbb{R}^{R\times R\times Q}$ is the coefficient tensor adjusting for the effects from the
covariates, and we expect $\mathcal{M}$ to be semi-symmetric under frontal slices given each $\mathcal{M}_{:,:,q}$ is symmetric. Similar to the way structural connectivity  impacting the outcome, we anticipate the influence from exposure to connectivity also functions through subgraphs. Specifically, under $H$ impacted network configurations, we assume the subgraph structure of configuration $h$ is captured by $\bfalpha_h\in \mathbb{R}^{R\times 1}$ under an overall effect $\eta_h$. Similarly, imposing sparsity on $\bfalpha_h$ allows us to capture each clique subgraph impacted by the exposure.  Since the model formulation for $\boldsymbol{G}_i$ only guarantees a symmetric structure, to ultimately link $\boldsymbol{G}_i$ to the hollow matrix $\bfA_i$, we will deduct the diagonal elements from $\boldsymbol{G}_i$ denoted by $\mbox{Dg}[\boldsymbol{G}_i]$ while keeping the off-diagonal elements fixed. For the residual errors, we have $\bfE_i=(e_{ikl})\in  \mathbb{R}^{R\times R}$ with $e_{ikl}\sim \mbox{N}(0,\sigma_1^2/2)$. By performing an operation of $\bfE_i-\bfE^T_i$, we zero out the diagonal elements to create a symmetric hollow matrix with a residual error variance to be $\sigma_1^2$ for each structural connection.

It is worth emphasizing that in practice, different types of brain connectivity preserve distinct biological architectures. For example, brain functional connectivity where connections are defined by statistical dependence among neuronal interactions presents a modular structure over the whole brain to calibrate different functional systems \citep{schwarz2008community,power2011functional}. We focus on structural connectivity here by modeling the subgraph structures along effect pathways in \eqref{eq:aft} and \eqref{eq:matrix}, consistent with the densely connected white matter fiber tracks within groups of nodes. Our approach is in sharp contrast to \cite{zhao2022bayesian} where functional connectivity represented by modular structure serves as the mediator for a continuous outcome. Additionally, though we develop the above model under a mediation framework, neither \eqref{eq:aft} nor \eqref{eq:matrix} has been considered before in a network-variate association analysis, and they can readily to be applied separately as needed in other contexts where the need for network-variate regression arises.

\subsection{Effect pathways}
To quantify the effect pathways among the survival outcome, exposure and network-variate mediator, we introduce counterfactual representations to define the effect measures under structural modeling framework \eqref{eq:aft} and \eqref{eq:matrix},. Let $\bfA(Z)$ be the value of the connectivity mediator under exposure $Z$ with $Z=z$ or $Z=z^*$ denoting the two states. Let $T(Z, \bfA(Z))$ be the counterfactual survival time under $Z$ and mediator $\bfA(Z)$. Given the covariates,  we define the natural indirect effect (NIE), natural direct effect (NDE) and total effect (TE) as
    \begin{equation}
\begin{aligned}\label{eq:effect}
    \mbox{NIE}&=\mathbb{E}[\log\{T(z,\bfA(z)\}]-\mathbb{E}[\log\{T(z,\bfA(z^*)\}];\\
    \mbox{NDE}&=\mathbb{E}[\log\{T(z,\bfA(z^*)\}]-\mathbb{E}[\log\{T(z^*,\bfA(z^*)\}];\\
    \mbox{TE}&=\mbox{NIE}+\mbox{NDE}=\mathbb{E}[\log\{T(z,\bfA(z)\}]-\mathbb{E}[\log\{T(z^*,\bfA(z^*)\}]. 
\end{aligned}
    \end{equation}
Following general practice for time-to-event outcomes \citep{vanderweele2011causal}, model \eqref{eq:effect} defines each effect component on the survival function scale, which in our case is the log-survival function aligning with the AFT model. In our specific application, NIE quantifies the expected change in log-survival time when the connectivity mediator changes from $\bfA(z^*)$ to $\bfA(z)$ with fixed genetic exposure; NDE characterizes the expected change in  log-survival time by alternating the genetic exposure when brain connectivity is fixed; and TE is the sum of NIE and NDE, and hence captures the overall change in log-survival time under different states of genetic exposure. We have the following proposition for each effect component.

\begin{figure}[H]
\centering
  \includegraphics[width=1\textwidth]{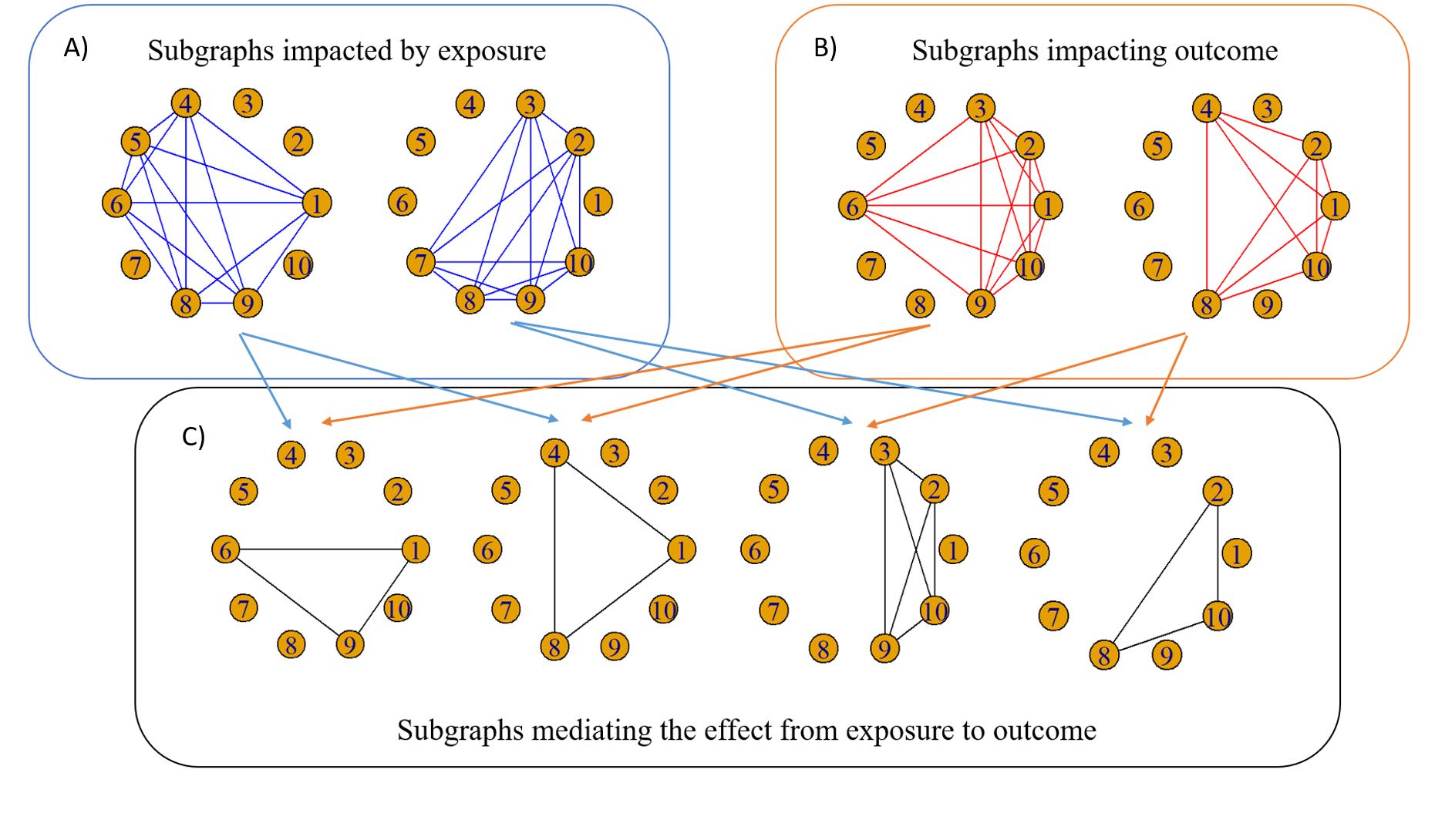}
  \vspace{-0.5cm}
    \caption{As a demonstration, the upper panels show  A) two informative subgraphs from the exposure to mediator effect pathway, and B) two informative subgraphs from the mediator to survival outcome effect pathway. The bottom panel provides C) the truly active mediating components contributing to the Natural Indirect Effect (NIE), which consist of the overlapping network configuration.
}\label{fig:demo}
\end{figure}

\bigskip
\noindent \textbf{Proposition 1.} \textit{Under the proposed joint model presented in \eqref{eq:aft} and \eqref{eq:matrix} and definitions of the different effect components as presented in \eqref{eq:effect},  it can be shown that }
    \begin{equation}
\begin{aligned}\label{eq:effect2}
     \mbox{NIE}&
     =(z-z^*)\sum_j\sum_h\Big\langle\omega_j\bfbeta_j\bfbeta_j^T, (\eta_h\bfalpha_h\bfalpha_h^T-\eta_h\mbox{Dg}[\bfalpha_h\bfalpha_h^T])\Big\rangle_{F}\\
      \mbox{NDE}&=\beta_z(z-z^*)\\
    \mbox{TE}
    &=(z-z^*)\sum_j\sum_h\Big\langle\omega_j\bfbeta_j\bfbeta_j^T, (\eta_h\bfalpha_h\bfalpha_h^T-\eta_h\mbox{Dg}[\bfalpha_h\bfalpha_h^T])\Big\rangle_{F}+\beta_z(z-z^*).
\end{aligned}
    \end{equation}
The detailed derivations for the above proposition are provided in Web Appendix A. Compared with the effect measures under an existing mediation analysis with a scalar or vector mediator, the above proposition reveals that the structural connectivity mediator contributes to the TE/NIE by combining subgraphs identified along the exposure to mediator effect pathway, and those identified along the mediator to outcome effect pathway. Moreover, we conclude from \eqref{eq:effect2} that there are subgraphs that absorb impact from the exposure without changing the outcome; and subgraphs that alter the outcome intrinsically without interacting with the exposure. To further elaborate, Figure \ref{fig:demo} illustrates the above statement using a demonstration example of  two informative subgraphs from the exposure to mediator pathway, and two from the mediator to survival outcome pathway. Eventually, the truly active mediators are the overlapping configurations that impact both pathways as shown in the lower panel of Figure \ref{fig:demo}. It is also worth noting that another commonly adopted way to define effects for survival data is with log-expected survival times as $\widetilde{\mbox{NIE}}=\log\{\mathbb{E}[T(z,\bfA(z)]\}-\log\{\mathbb{E}[T(z,\bfA(z^*)]\}$, $\widetilde{\mbox{NDE}}=\log\{\mathbb{E}[T(z,\bfA(z^*)]\}-\log\{\mathbb{E}[T(z^*,\bfA(z^*)]\}$ and $\widetilde{\mbox{TE}}=\widetilde{\mbox{NIE}}+\widetilde{\mbox{NDE}}$. Under our modeling framework, it turns out that the ultimate formulations for each effect metric are identical under these two definitions. In other words, the estimated NIE and NDE from our model can be interpreted in two ways. We also provide the derivations based on these definitions in  Web Appendix A.
    
Under a classical casual mediation framework, a series of assumptions \citep{imai2010general} are typically required to claim a causal effect. Specific to our current analytical framework, these assumptions include that there is no unmeasured confounding for the relationships between the time to disease onset and genetic exposure, between structure connectivity and genetic exposure, and between AD survival and connectivity with and without controlling for the genotype. In addition, with a right-censored time-to-event outcome, we additionally assume covariate-dependent censoring such that the censoring time is independent of the failure time given exposure and a set of covariates adjusted for in the AFT model. It is noteworthy that in practice, the above assumptions are frequently unverifiable, but some of the assumptions may be relaxed under specific scenarios \citep{chen2018high}. Hence, in existing neuroimaging applications \citep{lindquist2012functional,zhao2021multimodal} as well as the current study, mediation analysis is considered as an exploratory strategy over a rigorous dissection of causality, and sensitivity analysis for our specific setting merits additional research.

\section{Bayesian Inference}\label{sec:bayesian}

\subsection{Prior specification}\label{sec:prior}
We propose a fully Bayesian algorithm to estimate the model parameters. Compared with a mediation analysis under a frequentist paradigm, Bayesian modeling is capable of directly providing inference for each of the effect components, which is arguably more favorable for complex modeling frameworks. As mentioned previously,  extracting informative brain subgraphs for each path captured by $\{\bfbeta_j, \omega_j\}^J_{j=1}$ and $\{\bfalpha_h, \eta_h\}^H_{h=1}$ is as important as estimating the NIE, NDE and TE when uncovering the effect pathways. Denoting $\bfbeta_j=(\beta_{j1},\dots,\beta_{jR})$ and $\bfalpha_h=(\alpha_{h1},\dots,\alpha_{hR})$,  we assign point mass mixture priors for each element of $\bfbeta_j$ and $\bfalpha_h$ to impose sparsity as
\begin{align}\label{eq:ssvs}
    \beta_{jr}\sim (1-\gamma_{jr})\delta_0+\gamma_{jr}\mbox{N}(0,\upsilon_1);  \quad   \alpha_{hr}\sim (1-\tau_{hr})\delta_0+\tau_{hr}\mbox{N}(0,\upsilon_2),
\end{align}
with $j=1,\dots,J; h=1,\dots,H; r=1,\dots,R$. Here, $\delta_0$ represents a point mass at zero, $\upsilon_1$ and $\upsilon_2$ are variance components with a large value, and $\gamma_{jr}$ and $\tau_{hr}$ are the latent selection indicators introduced to identify nonzero elements within the two sets of network-variant coefficients. Specifically, in the case of $\gamma_{jr}=1$, we have $\beta_{jr} \neq 0$, indicating that the connections linked with node $r$ represent an informative subgraph impacting the time to disease onset. Similarly, when $\tau_{hr}=1$, we have $\alpha_{hr} \neq 0$ indicating that connections linked with node $r$ represent an informative subgraph influenced by the genetic exposure. 

To specify priors for latent indicators $\bfgamma=\{\gamma_{1r},\dots,\gamma_{Jr}\}_{r=1}^R$ and $\bftau=\{\tau_{1r},\dots,\tau_{Hr}\}_{r=1}^R$, one can either impose a non-informative Bernoulli distribution for each of the elements, or resort to a more informative prior by incorporating additional biological structure. Such structural information includes brain regions located symmetrically between right and left brain hemispheres, or networks coming from nodes that are connected at the population level. To first summarize the information, one could consider an indirected knowledge graph $\mathcal{G}_0$ by including an edge if two nodes are symmetric on hemispheres or 
connected populationally, and then  encourage smoothness on selection over $\mathcal{G}_0$ using a Markov random field (MRF) prior \citep{sun2018knowledge} 
    \begin{equation}
\begin{aligned}\label{eq:MRF}
p(\bfgamma, \bftau)\propto\exp\Bigg\{\mu\sum_{r}\left(\sum_{j}\gamma_{jr}+\sum_{h}\tau_{hr}\right)+\nu\sum_{r\sim r'}\left(\sum_j\gamma_{jr}\gamma_{jr'}+\sum_h\tau_{hr}\tau_{hr'}\right)\Bigg\}
\end{aligned}
    \end{equation}
with $r\sim r'$ indicating a connection at $\mathcal{G}_0$. In equation \eqref{eq:MRF}, $\mu$ controls the number of nodes selected in a subgraph, and $\eta$ impacts the smoothness reflected by the degree of confidence to include both nodes in the informative subgraphs if they are connected at a prior level. When both $\mu=\nu=0$, equation \eqref{eq:MRF} 
degenerates to non-informative bernoulli distributions without prior structural information incorporated.
 $\{\omega_j\}^J_{j=1}$ and $\{\eta_h\}_{h=1}^H$ characterize graph-level effects for the $J$ subgraphs impacting the outcome and $H$ subgraphs influenced by the exposure. To distinguish the role each subgraph plays in the effect mechanism, and more importantly to shrink the effects associated with subgraphs that link loosely with both outcome and exposure, we impose Laplace priors  as
\begin{equation}\label{eq:lap}
\begin{aligned}
\omega_j &\sim  \mathcal{L}(\lambda_{\omega}); \quad \eta_h \sim \mathcal{L}(\lambda_{\eta}); \quad j=1,\dots, J; ~~h=1,\dots,H,
\end{aligned}
\end{equation}
with shrinkage parameters $\lambda_{\omega}$ and $\lambda_{\eta}$. This is critical in practice when  $J$ and $H$ are unknown because it allows us  to set conservative values and allow the model to determine  the numbers of subgraphs. 
For the remaining parameters, we set the following priors: $\bfbeta_x\sim \mbox{N}(\bfzero,\mbox{I}\sigma_x^2)$ and $\beta_z\sim N(0, \sigma_z^2)$; 
we assume the coefficient tensor $\mathcal{M}$, which associates the covariates with the connectivity mediator, follows a symmetric rank-$K$ tensor decomposition as $\mathcal{M}=\sum_{k\in[K]}\bfa_{1k}\circ\bfa_{1k}\circ\bfa_{2k}$ with $\bfa_{1k}\in \mathbb{R}^R, \bfa_{2k}\in \mathbb{R}^P$ to downsize the number of parameters and 
$\bfa_{1k}\sim \mbox{N}(\bfzero,\mbox{I}\sigma_a^2)$ and $\bfa_{2k}\sim \mbox{N}(\bfzero,\mbox{I}\sigma_a^2)$ for $k=1,\dots,K$;  we use a non-informative  inverse gamma (IG) prior for variances $\sigma_z^2, \sigma_e^2, \sigma_0^2, \sigma_1^2$ with shape and scale parameters equaling 0.01; and we use a large value (i.e., 10) for the remaining of  variance parameters.

\begin{figure}
    \centering
    \includegraphics[width=0.80\textwidth]{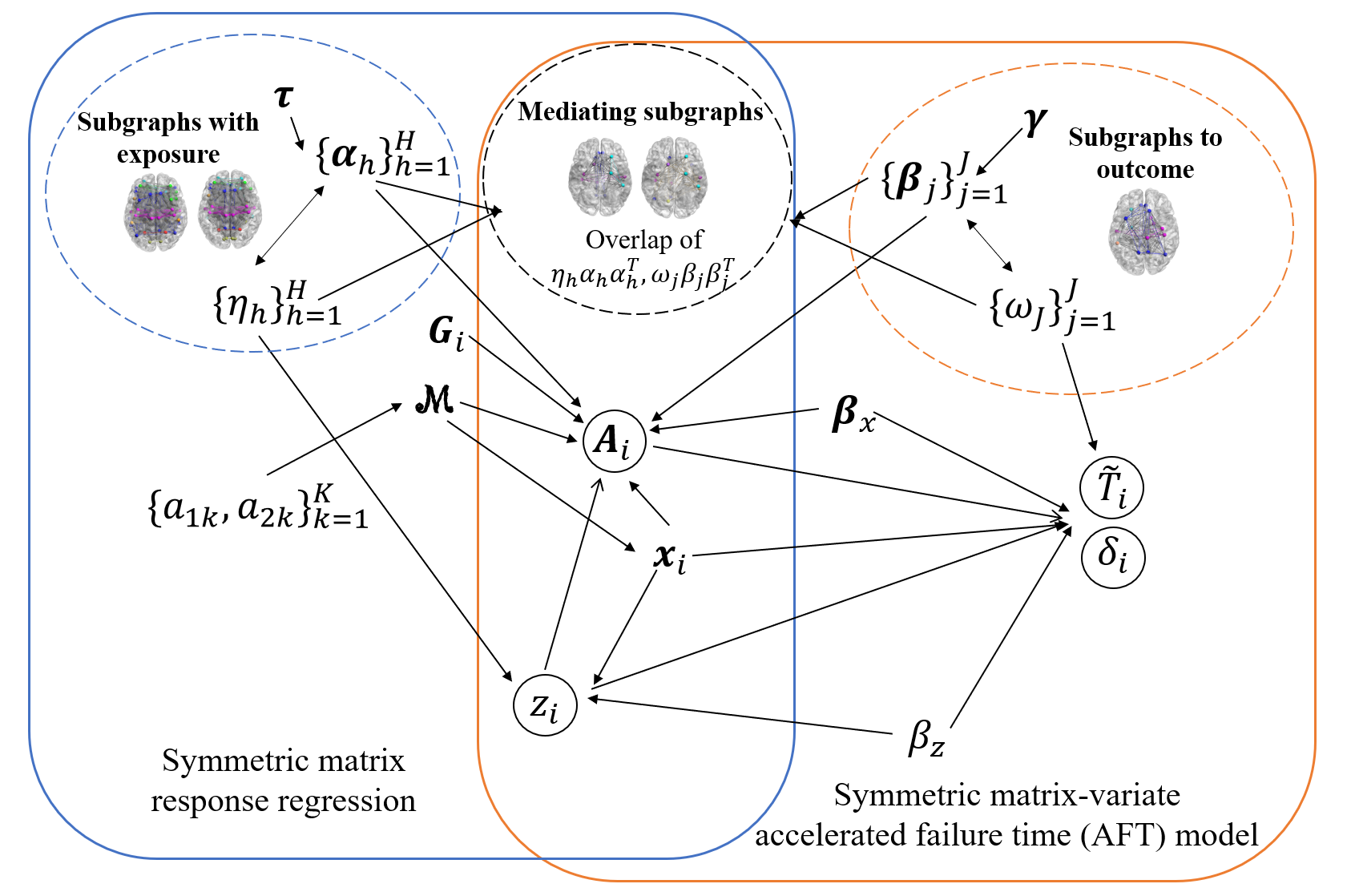}
    \caption{An illustration of the proposed  \textbf{B}ayesian \textbf{S}u\textbf{G}raph-based \textbf{M}ediation (BSGM) modeling framework for identifying and characterizing effect components induced by brain subgraphs along with the input and output data structure. In our structural modeling, we develop an innovative symmetric matrix response regression and an innovative symmetric matrix-variate accelerated faiure time (AFT) model with the data and major parameters involved in each component provided.  }.
    \label{fig:demo0}
\end{figure}

Finally, there are a few tuning parameters needed to be pre-specified before implementing our model, including the upper bound of the number of informative subgraphs, $H$ and $J$,  decomposition rank $K$, and sparsity and smoothness parameters $\mu$ and $\nu$, respectively, when an MRF prior is used. Given $\mathcal{M}$ is a nuisance parameter and its estimation is not of interest, we directly specify $K$ to a be a reasonable number. In our  application we set $K$ to be 3 in order to capture sufficient information in $\mathcal{M}$. For the rest of the tuning parameters, we consider a grid of values, and implement our model under all  parameter combinations. Then we choose the optimal one using Bayesian information criterion (BIC). 
We name our model \textbf{B}ayesian \textbf{S}u\textbf{G}raph-based \textbf{M}ediation (BSGM) analysis, and present a demonstration of our analytical framework in Figure \ref{fig:demo0}.

\subsection{Posterior Inference}
Given the observed data $\mathcal{O}=(\widetilde{T_i}, \bfA_i, \bfx_i, z_i, \delta_i; i=1,\dots, N)$, the unknown parameters $\bfTheta=\bigg[\bfbeta_x, \beta_z, \{\bfalpha_h, \eta_h\}_{h=1}^H,\{\bfbeta_j, \omega_j\}_{j=1}^J,
\bftau, \bfgamma, \{\bfa_{1k}, \bfa_{2k}\}_{k=1}^K, \sigma_z^2, \sigma_e^2, \sigma_0^2, \sigma_1^2\bigg]$ follow the joint conditional posterior distribution: 
\begin{align*}
f(\bfTheta|\mathcal{O})\varpropto \prod_{i} &\bigg\{\delta_i f(\widetilde{T_i}\mid \bfTheta )+(1-\delta_i)S(\widetilde{T_i}\mid \bfTheta )\bigg\}
  \prod_{i}f(A_i \mid  \bfTheta )\times\\
\prod_{h}\prod_{j}& \bigg\{f(\eta_h)f(\omega_j)f(\bfalpha_h|\bftau_h)f(\bfbeta_j|\bfgamma_j)f(\bfgamma_j)f(\bftau_h)\bigg\}f(\bfbeta_x)f(\beta_z)f(\sigma_e^2)f(\sigma_0^2)f(\sigma_z^2)\times\\
\prod_{k}&\bigg\{f(\bfa_{1k})f(\bfa_{2k})\bigg\},
\end{align*}
where $f(\widetilde{T_i}\mid \bfTheta)$ and  $S(\widetilde{T_i}\mid \bfTheta)$ are the conditional density and survival functions, respectively. 
Based on the assumption that the time-to-event follows a log-normal distribution, the survival function $S(\widetilde{T_i}\mid \bfTheta)=1-\bfPhi\left(\frac{\log(\widetilde{T_i})-y_i}{\sigma_0}|\Theta\right)$, where $\bfPhi(\cdot)$ is the cumulative normal distribution, $\sigma_0$ is the variance of the residual term of the AFT model, and $y_i$ is the regression term of the AFT model. 
We develop a Markov chain Monte Carlo (MCMC) algorithm through Gibbs samplers to conduct posterior inference. A brief overview of the sampling steps is shown below with the detailed sampling procedure included in Web Appendix B. 

\begin{itemize}
\item For the symmetric bilinear AFT model, update the covariates and exposure coefficients $\bfbeta_x$ and  $\beta_z$ from their corresponding posterior multivariate Normal distributions.
\item For the symmetric matrix response regression, update the covariates' coefficients tensor $\mathcal{M}$ by sampling each element within $\{\bfa_{1k}, \bfa_{2k}\}_{k=1}^K$ from the corresponding posterior Normal distributions. 
\item For the shrinkage parameters $\{\omega_j\}$ and $\{\eta_h\}$, we represent each of their Laplace priors by a scale mixture of Normals with an exponential mixing density. This allows us to derive a hierarchical representation of the prior model and update hierarchically from full conditionals \citep{park2008bayesian} for each of the elements.
\item For subgraph related coefficients $\{\bfalpha_h\}$ and $\{\bfbeta_j\}$, update each of these coefficient vectors from their corresponding posterior Normal distributions given the selection indicators $\bftau$ and $\bfgamma$, respectively.
\item For the selection indicator vectors $\bftau, \bfgamma$, update each element from the corresponding posterior Bernoulli distributions.
\item Update each $\sigma_z^2, \sigma_e^2, \sigma_0^2, \sigma_1^2$ from theits corresponding posterior IG distribution.

\end{itemize}

After obtaining the posterior samples from the above MCMC algorithm, based on the marginal posterior probability for the elements in $\bfgamma$ and $\bftau$, and the posterior mean for each $\{\eta_h\}$ and $\{\omega_h\}$, we are able to identify the informative brain subgraph configurations along each effect pathway. The informative connections within the subgraph are then determined based on the posterior probability of $\bfgamma$ and $\bftau$ using a cutoff value of 0.5  in light of the median probability model \citep{barbieri2004optimal}. Subsequently, the posterior estimates for the NIE, NDE and TE and their credible intervals can  be obtained directly from \eqref{eq:effect2}.

\bigskip

\section{Simulation Study}\label{sec:simu}

We carry out simulation studies to evaluate the performance of the proposed mediation analysis and compare the proposed model to  alternative modeling approaches. To mimic the data dimension in our application, we consider sample sizes of $N=100$ and $500$ with each subject's brain connectivity $R=100$. For the informative network configurations linked with the exposure and outcome, to assess the robustness of our method under unfavorable cases, we first generate three subgraphs for the effect pathway to outcome and then three separate subgraphs for the pathway from the exposure. We set $\eta=(0.8, 0, 0.9)$ and $\omega=(1.2, 0, 0)$, allowing some of the subgraphs to be noisy without contributing to the effect mechanism. In other words, we have two subgraphs influenced by the exposure and one subgraph impacting the outcome, and the true mediating network configurations are their overlapping components. For the effect coefficients $\bfalpha_h$ and $\bfbeta_j$ ($h,j=1, 2, 3$) within each subgraph, we assume 15\% of the elements are nonzero corresponding to the clique signaling graphs with the value of each coefficient vector shown in Web Figure 1.  To examine the performance of our model when the subgraph coefficients do not need to be shrunk, we also consider a simple case by generating one subgraph for the effect pathway to the outcome, and one subgraph for
the pathway from the exposure. Similarly, we set $\eta=0.8$, $\omega=1.2$, and assume 15\% of the elements are nonzero for $\bfalpha_h$ and $\bfbeta_j$ ($h,j=1$). For the exposure, we generate $z_i\sim \mbox{Bern}(0.5)$ from a Bernoulli distribution and consider three covariates in addition to the intercept term with the first two covariates generated from a standard Normal distribution and the last  generated from a Bernoulli distribution with probability 0.65. We set $\beta_z$ to be 1.4, $\bfbeta_x=(1.4,0.8,0.5)$ and generate the coefficient tensor $\mathcal{M}$ based on a rank-3 symmetric decomposition with $\bfa_{1k}\sim \mbox{N}(0, 0.4), \bfa_{2k} \sim \mbox{N}(0, 0.4)$ for $k=1,2,3$. For the residual error variances, we assume $\sigma^2_0=\sigma^2_1=0.4$. To assess the model fitting under different censoring schemes, we generate the censoring time $R_i$  under two scenarios: \textbf{A}) independent censoring with $R_i\sim \mathcal{E}(0.4)$ where $\mathcal{E}(\cdot)$ denotes an Exponential distribution; and \textbf{B}) covariate-dependent censoring with $R_i\sim \mathcal{E}(\xi^T\bf_i)$, where $\xi=(0.5,0.2,0.4)$. After obtaining $\widetilde{T}_i$, we directly determine the censoring indicator $\delta_i$. Based on the above design, we calculate the true effect measures (NIE, NDE and TE). In total, we consider eight  simulated settings, and we generate 100 Monte Carlo datasets for each setting.

To implement the proposed BSGM,  we adopt non-informative Bernoulli priors for $\bfgamma$ and $\bftau$ with no prior structural information incorporated. We set both $H$ and $J$ to be 3, which is larger than the actual number of informative subgraphs in each effect pathway, and we specify $K$ to be 3. The MCMC algorithm is conducted under random initials for 10,000 iterations with 5,000 burn-in.  The posterior convergence is confirmed using both trace plots and the Gelman-Rubin method \citep{gelman1992inference}. Given none of the existing methods can directly model a network mediator with a survival outcome, we first consider existing univariate or multivariate mediator analyses as comparisons. Specifically, we vectorize the upper diagonal elements of each $\bfA_i$ to extract the unique connections and include each connection as a mediator for univariate-mediator mediation analyses under the {\tt{R}} package {\tt{mediation}} (UMA), as well as all of them jointly as one mediator vector for multivariate-mediator mediating analysis implemented by the {\tt{R}} package {\tt{mma}} (MMA). In addition, we consider the Bayesian network mediation model (BNMM) \citep{zhao2022bayesian}, which is the only method to the best of our knowledge, that can handle a network mediator. Given the BNMM is designed for continuous outcomes, we have to adjust its model formulation to make it feasible for survival outcomes. Finally, to evaluate the performance of each method, we focus on the accuracy of estimating the effect elements and identifying the mediating components. These evaluation metrics include the mean or posterior mean (\textit{Mean}), percentage bias (\textit{Bias}), and coverage (\textit{Coverage}) by confidence interval (UMA, MMA) or credible interval (BSGM, BNMM) for estimating NIE, NDE and TE, and the \textit{Sensitivity} and \textit{Specificity} for selecting the truly active mediators. Of note, to maintain consistency when evaluating selection accuracy, the identified mediating subgraphs or edges are always mapped back to the same input connectivity matrix before calculating sensitivity and specificity when compared with the ground truth.  The simulation results are summarized in Table \ref{table:result}.

Based on the simulation results, we conclude that our proposed BSGM outperforms the competing methods in uncovering effect mechanisms  under all the settings with different sample sizes, subgraph configurations and censoring schemes. To elaborate,  BSGM achieves a small percentage bias, close to 95\% coverage for all the effect components (NIE, NDE and TE), and close to 100\% sensitivity and specificity for identifying the truly active mediating subgraghs in all the simulated settings. These results indicate the strong power of our method to uncover each effect pathway and its related brain network signals.  Even though we implemented the BSGM under $H$ and $J$ larger than the true number of informative subgraphs in each effect pathway, our model still correctly uncovered them in general, reflecting the robustness of our method under reasonably set $H$ and $J$. For the competing methods, across all the available settings, UMA and MMA are inferior, with much larger estimation biases for different effects and more conservative identification of mediators compared with our method. This is anticipated given the network configurations functioning along each effect pathway are completely eliminated in these methods, indicating the necessity to exploit the network structure for network-variate mediators.  The BNMM, though considering a network mediator as a whole unit before decomposing it via stochastic block structures, fails to reflect the subgraph configurations we expect for connectivity, leading to poor estimation and identification accuracy under our current simulated settings. 

\begin{table}
\centering
 \caption{Simulation results under different sample sizes, censoring scenarios A) independent censoring  and B) covariate-dependent censoring, and number of subgraphs $H=J=1; H=J=3$. The evaluation metrics include mean or posterior mean (Mean), percentage bias (Bias), and coverage (Coverage) for the Natural Indirect Effect (NIE), Natural Direct Effect (NDE) and Total Effect (TE), and the sensitivity and specificity for identifying mediating connections in the network.}\label{table:result}
 \vspace{0.2cm}
\scalebox{0.65}{
\begin{tabular}{ccccrrrrrrr}
\toprule
                &     &     &&         \multicolumn{4}{c}{$N=100$}   &      \multicolumn{3}{c}{$N=500^*$} \\
                \cmidrule(lr){5-8} \cmidrule(lr){9-11}
{Number of Subgraphs} & Scenario& & &BSGM& UMA&MMA&BNMM&BSGM& UMA&BNMM  \\
\hline
  && &NIE&6.80&-0.01&-3.43&9.02&6.75&0.01&8.31\\
&&Mean&NDE&1.28&20.65&3.56&0.23&1.38&5.80&0.57\\
&&&TE&8.28&20.64&0.13&9.25&8.13&5.81&8.89\\
\cmidrule(lr){5-8} \cmidrule(lr){9-11}
 &&&NIE&1.19&-100.72&-151.04&10.40&0.45&-100.00&-13.41\\
 && Bias (\%)&NDE&-8.57&1375.00&154.29&-83.57&-1.42&314.29&-12.85\\
 &\textbf{A}&&TE&1.97&154.19&-98.39&-3.34&0.12&-39.18&-21.31\\
 \cmidrule(lr){5-8}\cmidrule(lr){9-11}
 &&&NIE&92.86&1.15&5.24&62.08&92.55&0.00&34.52\\
 &&Coverage (\%)&NDE&96.81&1.41&9.88&54.94&86.94&91.12&42.92\\
 &&&TE&97.58&81.41&1.65&59.65&91.67&95.68&52.08\\
 \cmidrule(lr){5-8}\cmidrule(lr){9-11}
 &&Sensitivity(\%)&&0.95&0.10&0.00&0.00&1.00&0.20&0.30\\
\textbf{1}&&Specificity(\%)&&0.95&0.98&0.99&0.99&0.95&0.99&0.90\\
 \cmidrule(lr){2-11}
&&&NIE&6.78&0.03&1.64&1.98&6.85&0.01&3.53\\
 &&Mean&NDE&1.25&4.95&-2.25&1.14&1.24&7.58&2.25\\
 &&&TE&8.03&4.98&-0.61&3.12&8.09&7.59&5.78\\
 \cmidrule(lr){5-8}\cmidrule(lr){9-11}
 &&&NIE&0.89&-100.71&-75.80&-70.54&1.93&-100.00&-47.47\\
 &&Bias (\%)&NDE&-10.71&253.57&365.00&-18.57&-11.42&441.42&60.71\\
 &\textbf{B}&&TE&-1.11&-38.66&-107.51& -61.57&-0.36&-6.52&-28.82\\
 \cmidrule(lr){5-8}\cmidrule(lr){9-11}
  &&&NIE&100.00&0.07&0.00&10.56&95.00&0.00&32.68\\
 &&Coverage (\%)&NDE&91.25&99.97&1.02&59.86&89.57&0.70&16.93\\
 &&&TE&98.72&99.98&1.02&14.72&95.96&99.95&33.42\\
 \cmidrule(lr){5-8}\cmidrule(lr){9-11}
 &&Sensitivity(\%)&&1,00&0.40&0.10&0.20&1.00&0.00&0.10\\
 &&Specificity(\%)&&0.95&0.99&0.96&0.96&0.90&0.99&0.95\\
 \hline
 & & &NIE&8.36&0.01&-2.71&16.71& 8.30&0.02&3.46\\
&&Mean&NDE&1.18&11.96&4.32&1.45&1.25&10.03&0.99\\
&&&TE&9.54&11.97&1.61&18.16&9.54&10.05&4.45\\
\cmidrule(lr){5-8} \cmidrule(lr){9-11}
 &&&NIE&2.32&-100.81&-133.17&104.53&1.63&-100.00&-57.65\\
 && Bias (\%)&NDE&16.13&754.50&208.57&4.57&-11.46&646.53&-29.29\\
 &\textbf{A}&&TE&1.02&25.33&-83.12&89.76&1.03&5.79&-53.50\\
 \cmidrule(lr){5-8}\cmidrule(lr){9-11}
 &&&NIE&93.62&0.01&16.84&10.57&92.52&0.22&23.62\\
 &&Coverage (\%)&NDE&89.36&1.00&10.12&80.34&94.43&0.20&35.68\\
 &&&TE&99.00&98.99&4.84&12.21&100.00&100.00&30.80\\
 \cmidrule(lr){5-8}\cmidrule(lr){9-11}
 &&Sensitivity(\%)&&1.00&0.20&0.0.00&0.20&1.00&0.10&0.15\\
 \textbf{3}&&Specificity(\%)&&0.95&0.98&0.99&0.95&0.96&0.99&0.90\\
 \cmidrule(lr){2-11}
 &&&NIE&7.65&-0.01&1.22&5.24&8.10&0.01&7.56\\
 &&Mean&NDE&1.09&6.24&-1.49&0.01&1.67&9.02&-0.03\\
 &&&TE&8.74&6.23&-0.27&5.25&9.77&9.03&7.53\\
 \cmidrule(lr){5-8}\cmidrule(lr){9-11}
 &&&NIE&-6.12&-100.00&-85.07&-35.86&-0.24&-100.00&-7.47\\
 &&Bias (\%)&NDE&-22.86&345.71&-192.14& -100.71&19.29&544.28&-102.63\\
&\textbf{B}&&TE&-8.28&-34.80&-102.83&-45.14&2.52&-5.64&-12.14\\
 \cmidrule(lr){5-8}\cmidrule(lr){9-11}
 &&&NIE&88.65&0.53&3.12&40.28&91.86&0.04&64.96\\
&&Coverage (\%)&NDE&90.54&0.87&0.00&10.20&88.38&0.04&4.78\\
&&&TE&94.28&4.02&1.54&20.50&89.96&99.79&62.84\\
 \cmidrule(lr){5-8}\cmidrule(lr){9-11}
&&Sensitivity(\%)&&0.90&0.40&0.20&0.10&1.00&0.00&0.00\\
&&Specificity(\%)&&1.00&0.99&1.00& 0.85&0.99&0.99&0.90\\
     \bottomrule   
     \multicolumn{11}{l}{\small * Due to the number of poorly fitting observations, MMA fails to complete under N = 500 ($>200$ hours). } \\
\end{tabular}}
\end{table}

\section{Application to the Alzheimer’s Disease Neuroimaging Initiative study}\label{sec:ADNI}

We then apply our method to the data extracted from a landmark AD study, the Alzheimer’s Disease Neuroimaging Initiative (ADNI) database (\url{adni.loni.usc.edu}). The ADNI was launched in 2003 by the National Institute on Aging (NIA), the National Institute of Biomedical Imaging and Bioengineering (NIBIB), the Food and
Drug Administration (FDA), private pharmaceutical companies, and nonprofit organizations, as a public–private partnership. The study includes ADNI1, ADNI GO/2 and ADNI3, and its primary goal was to test whether serial magnetic resonance imaging
(MRI), positron emission tomography (PET), other biological
markers, and clinical and neuropsychological assessment can
be combined to measure the progression of mild cognitive impairment (MCI) and early AD \citep{ADNI2011}.  In our current application, we are interested in exploring the effect pathways among APOE $e4$ genotype, brain structural connectivity and time to AD onset right-censored by death or loss to follow-up. We focus on the 119 ADNI2 participants who have both baseline magnetic resonance imaging (MRI) and  diffusion tensor imaging (DTI) data collected in order to extract white matter fiber tracts to create structural connectivity. Among these participants, 26 experienced AD onset with the rest censored by death and loss of follow-up; and the APOE $e4$ genotype rate is 0.48. In our analyses, we also include gender, age at the screening visit, and years of education as covariates.

To construct structural connectivity for each subject, we first perform anatomical parcellation on the high-resolution T1-weighted anatomical MRI scan to obtain 68 gyral-based ROIs through the FreeSurfer. We employ the Lausanne parcellation scheme to subdivide these ROIs into 83 small ROIs. After pre-possessing, including correction for motion and eddy current effects in DTI images, the DTI data are output to Diffusion Toolkit for fiber tracking. The FACT (fiber assignment by continuous tracking) algorithm is performed \citep{DTI2017} to initialize tracks from many seed points. It propagates these tracks along the most significant principal axis vector within each voxel until specific termination criteria are met. In this application, we characterize each connection by the number of fiber tracks connecting the two ROIs divided by the mean volume of the corresponding ROI pairs. This produce the structural connectivity for each subject summarized by an $83\times 83$ connectivity matrix. As a demonstration, we randomly select a subject and show the fiber traits and the summarized connectivity matrix in Web Figure 2.

\begin{figure}
    \centering
    \includegraphics[width=1.05\textwidth]{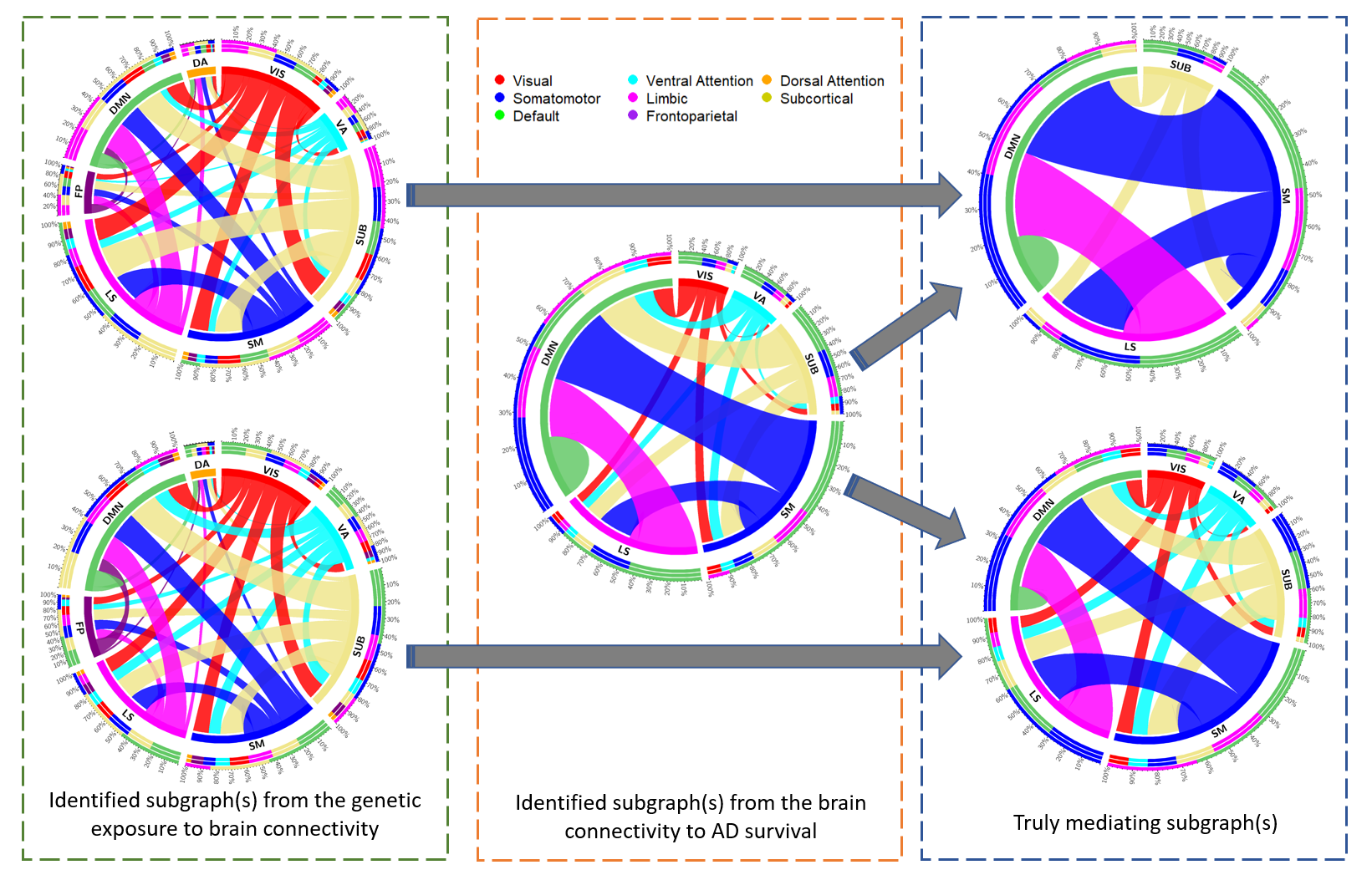}
    \caption{The brain network view for the identified subgraphs along each effect pathway as well as the truly mediating subgraphs. }
    \label{fig:my_label}
\end{figure}

We perform the desired mediation analysis using our proposed BSGM with implementation settings following those in the simulations. As a result, we identify two subgraphs influenced by the genotype and one subgraph impacting the outcome. The truly mediating subgraphs that bridge the genetic exposure to AD survival are obtained directly by overlapping these network components. Based on the posterior samples, we also estimate the NDE, NIE and TE with $95\%$ credible intervals as $-1.36$ $(-2.14, -0.68)$, $-0.04$ $(-0.55, 0.43)$ and $-1.41$ $(-2.44, -0.60)$, respectively. These results quantify the expected change on the log-survival time when altering genetic exposure, brain connectivity, and both. All three estimations are negative, indicating that APOE $e4$ will induce faster AD onset through both its direct effect, as well as indirect effect mediated by brain structural connectivity. Both the NDE and TE are significant with their 95\% credible intervals excluding zero. To further investigate how this effect mechanism  functions along the identified brain subgraphs, we map the nodes in each subgraph to the canonical functional systems \citep{power2011functional} summarized in Table \ref{table:real}. As can be seen, the majority of the identified subgraphs involve  Default Mode, Limbic and Somatomotor systems. Existing AD literature reveals that Default Mode Network is one of the most well-known neuroimaging biomarkers for AD \citep{default2016}. The Limbic system, which involves memory and emotion, is also severely and routinely affected during neurodegeneration \citep{limbic1976} with a recent study indicating that Limbic system is simultaneously affected by APOE $e4$ regarding the volume and shape of the amygdala and the hippocampus \citep{apoe2015}. We also provide visualizations for the identified subgraphs and the effect pathways they belong to  in Figure \ref{fig:my_label}, and illustrate the identified subgraphs are filled with cross-system connections. For the truly mediating subgraphs, most of the cross-system connections are also among the Default Mode, Limbic, and Somatomotor systems. In contrast to the cross-system connections, there are only a small proportion of within-system connections as shown in Figure \ref{fig:my_label}. This phenomenon is in accordance with previous AD literature  \citep{withinconnect} on sparse connections within functional systems in causing clinical symptoms and cognitive deficits.

\begin{table}[H]
\caption{Identified sub-networks within the mediation analyses and their overlaps with the functional system. }\label{table:real}
 \begin{center}
 \resizebox{0.9\textwidth}{!}{
  \begin{tabular}{cccccc}
  \toprule
   &  \multicolumn{2}{c}{Exposure} & \multicolumn{1}{c}{Outcome}& \multicolumn{2}{c}{Active} \\
   \cmidrule(lr){2-3}\cmidrule(lr){5-6}
 Functional System   & Subgraph 1& Subgraph 2 & Subgraph &Subgraph 1& Subgraph 2\\
 \hline
   Default Mode & 10 & 14& 7 &5&4\\
             
 Dorsal Attention & 2 & 2 & 0&0&0  \\
            
  Frontoparietal  & 3 & 5 & 0 &0&0\\
  Limbics  & 12 & 10 & 3 &3&3\\
 Somatomotor  & 10 & 11 & 4&4&4\\
Subcortical  & 12 & 12 & 2&1 &2\\
Ventral Attention  & 3 & 6 & 1&0&1 \\
Visual  & 8 & 8 & 1&0&1 \\
  \hline
   $\#$of Nodes&60&68&18&13&15\\
   Weight Mean&0.57&0.73&0.21&0.11&0.14\\
  
\bottomrule
  \end{tabular}
  }
  \end{center}

\end{table}

\begin{figure}[H]
\centering
\captionof*{figure}{Positive and negative networks within truly mediating subgraph 1}
  \includegraphics[width=.24\textwidth]{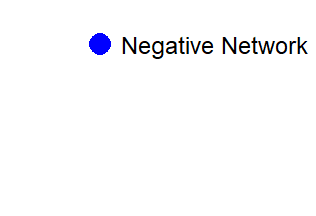}
    \includegraphics[width=.24\textwidth]{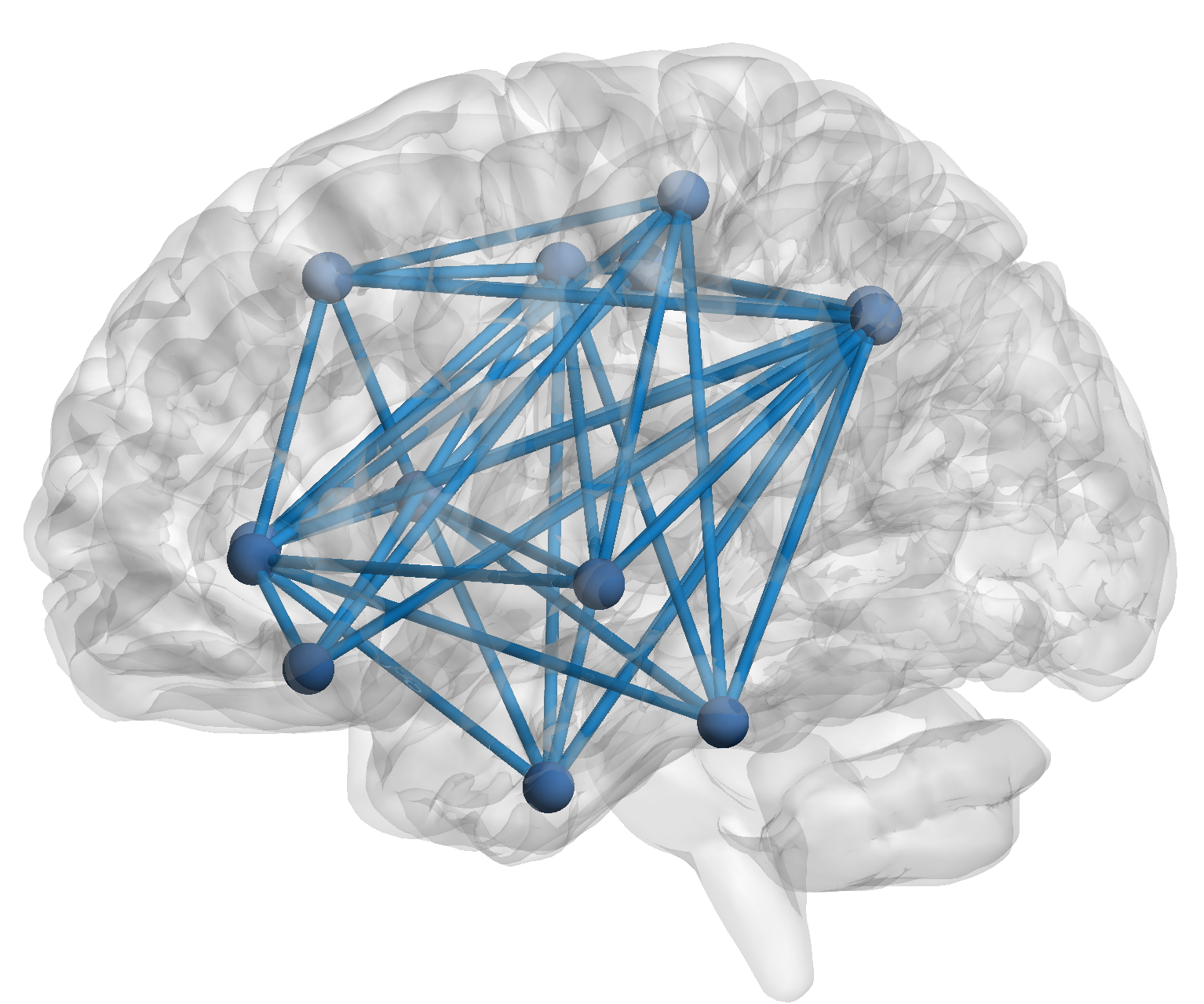}
     \includegraphics[width=.24\textwidth]{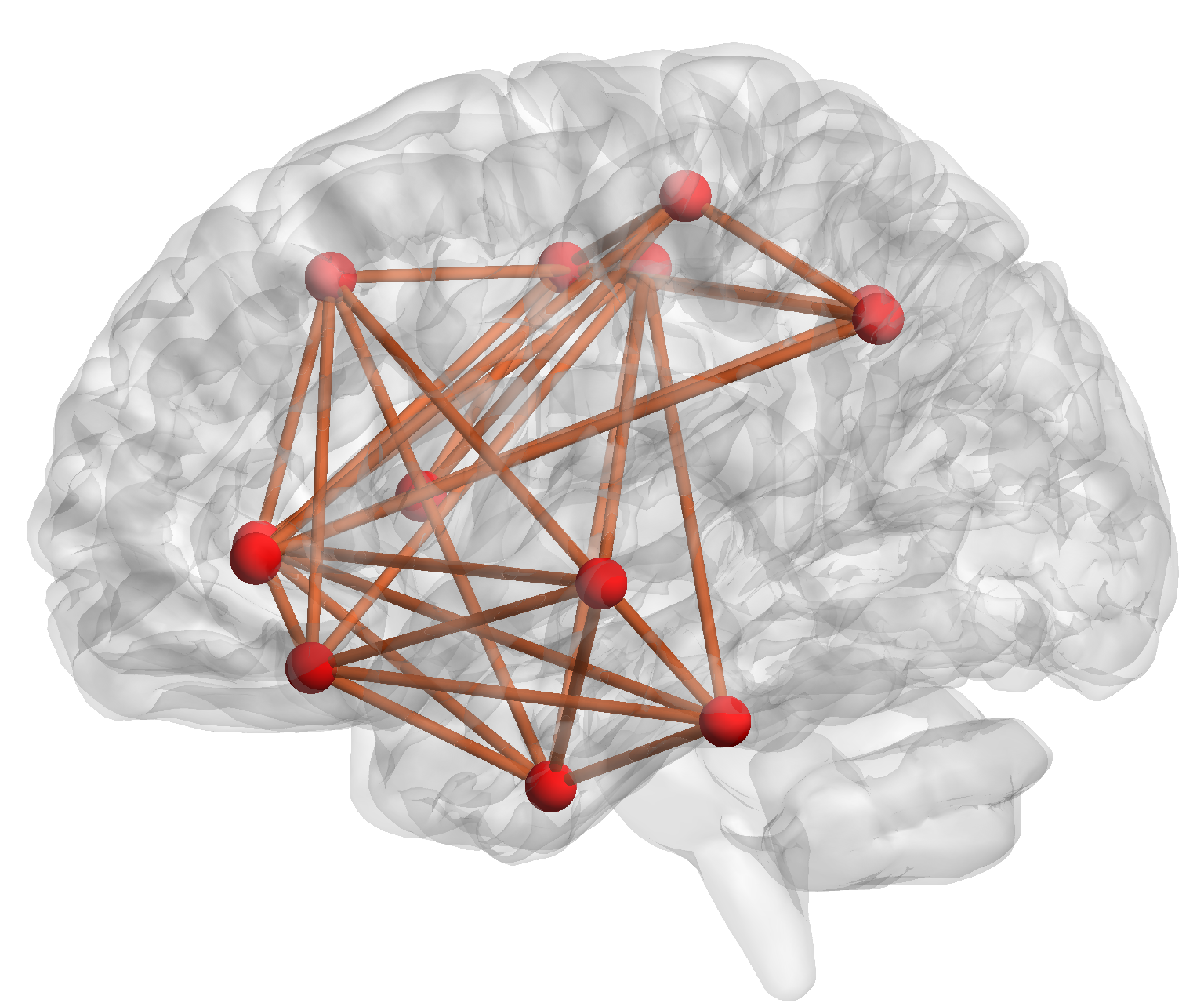}
      \includegraphics[width=.24\textwidth]{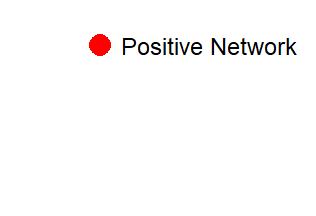}
      
    \includegraphics[width=.24\textwidth]{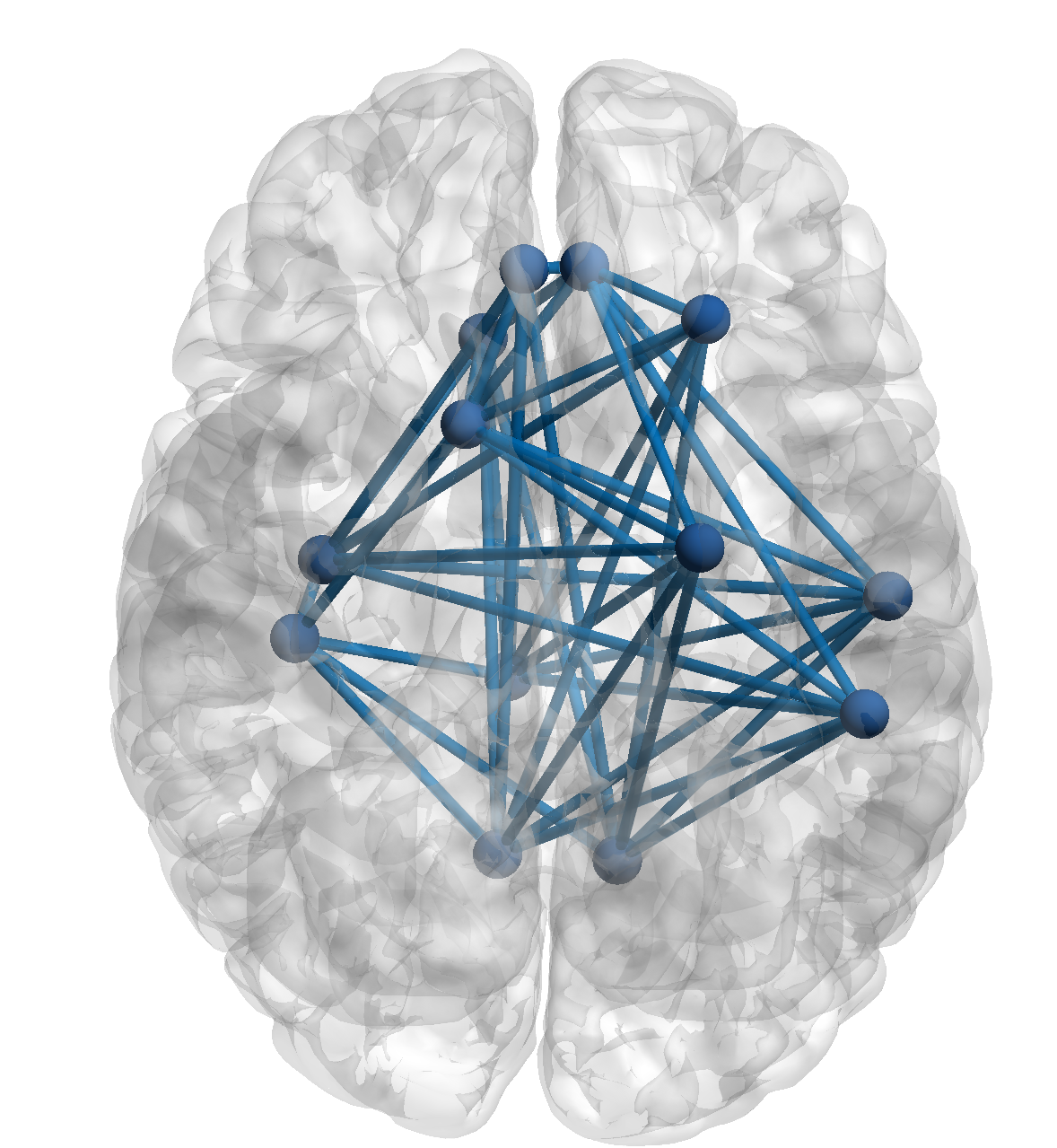}
    \includegraphics[width=.24\textwidth]{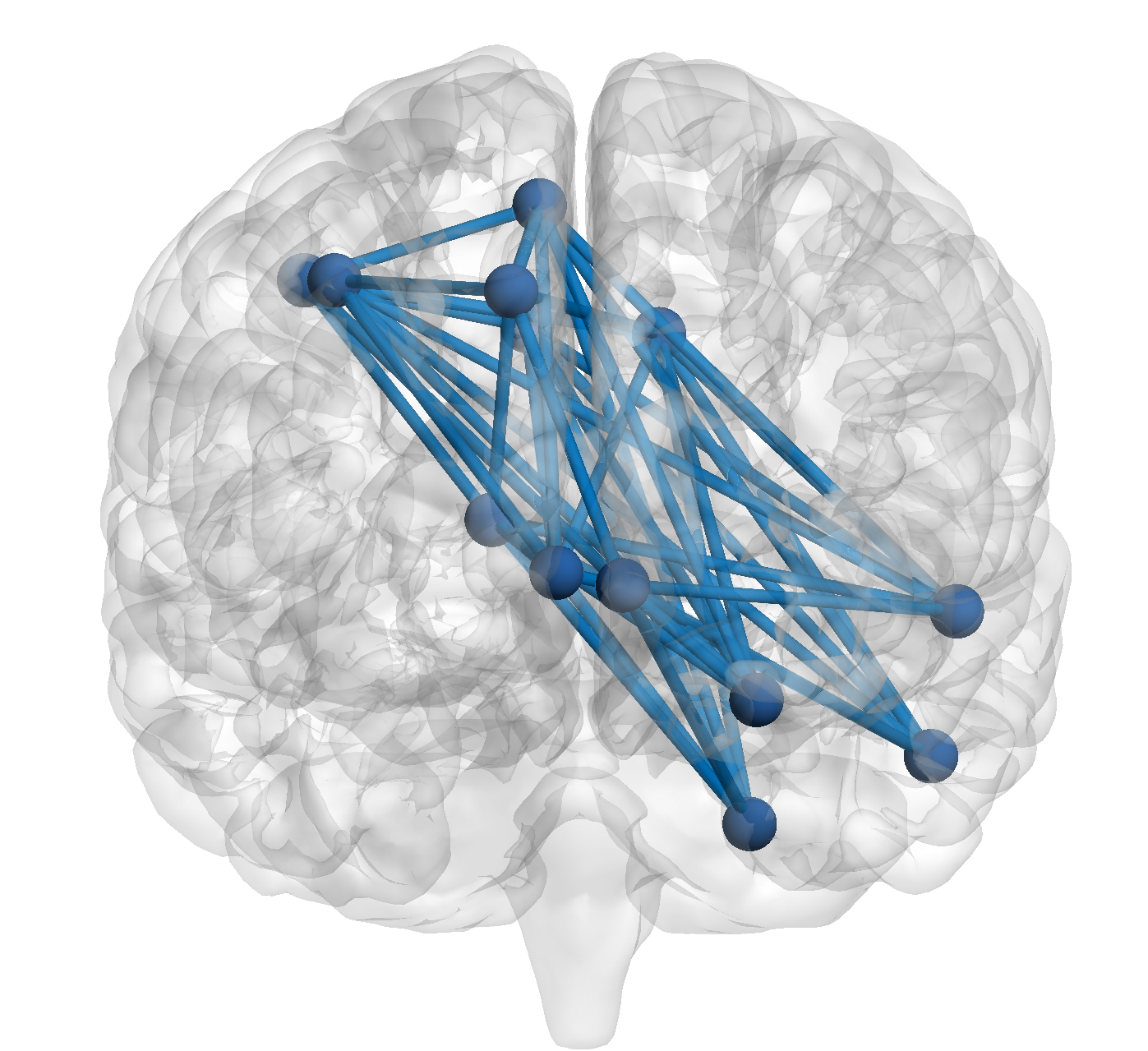}
     \includegraphics[width=.24\textwidth]{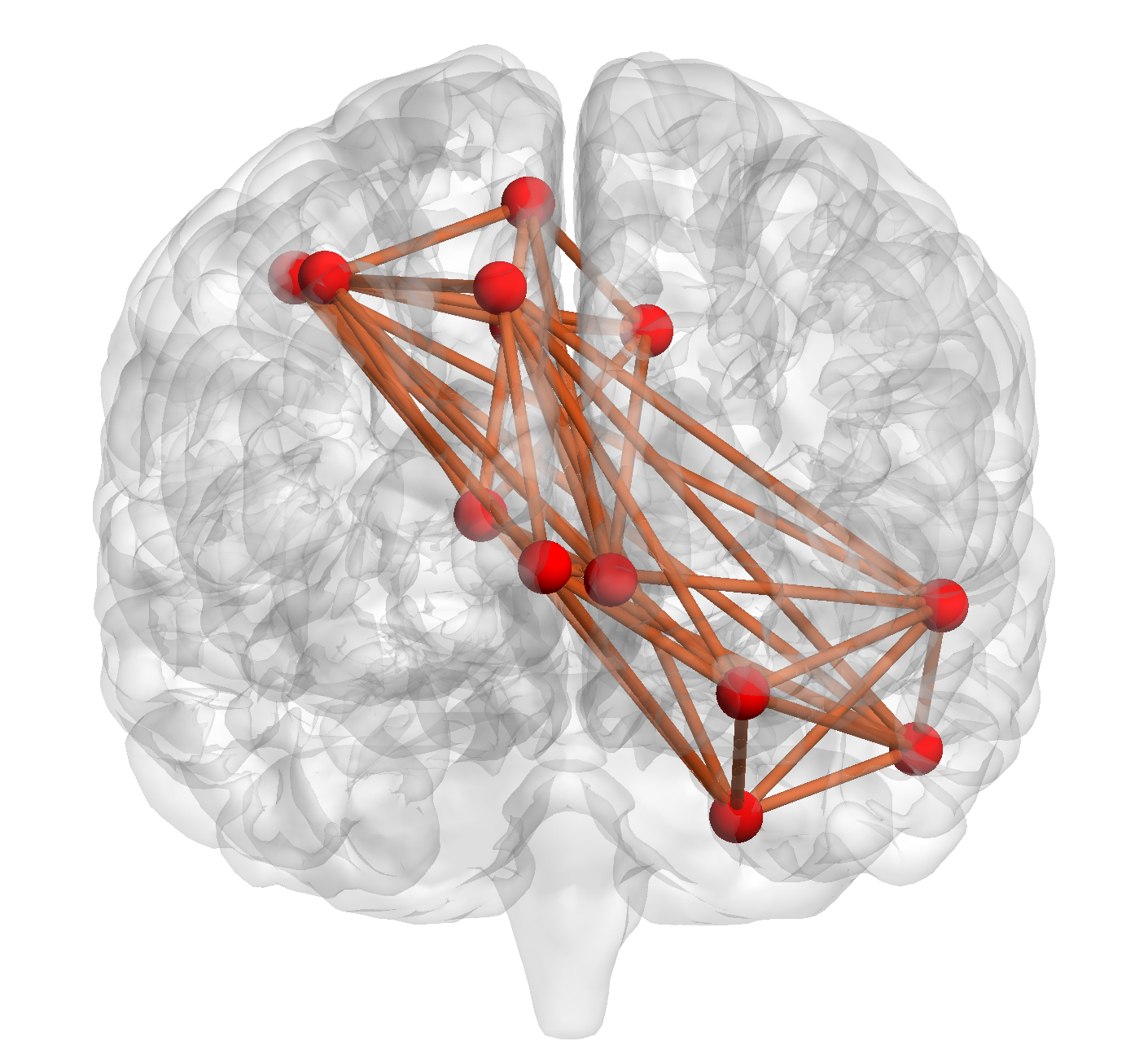}
     \includegraphics[width=.24\textwidth]{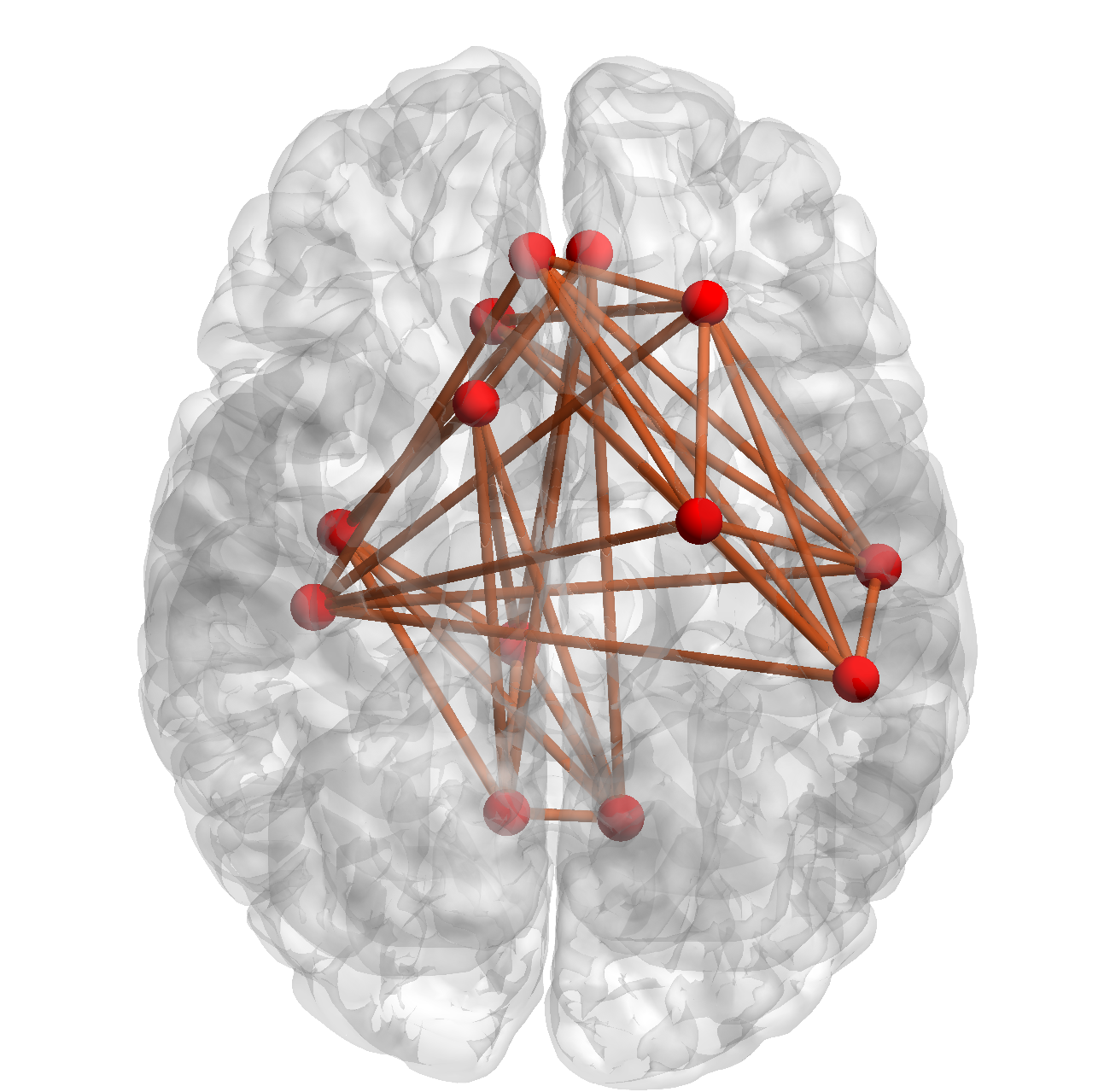}

     \captionof*{figure}{Positive and negative networks within truly mediating subgraph 2}
  \includegraphics[width=.24\textwidth]{figure/legend1.png}
    \includegraphics[width=.24\textwidth]{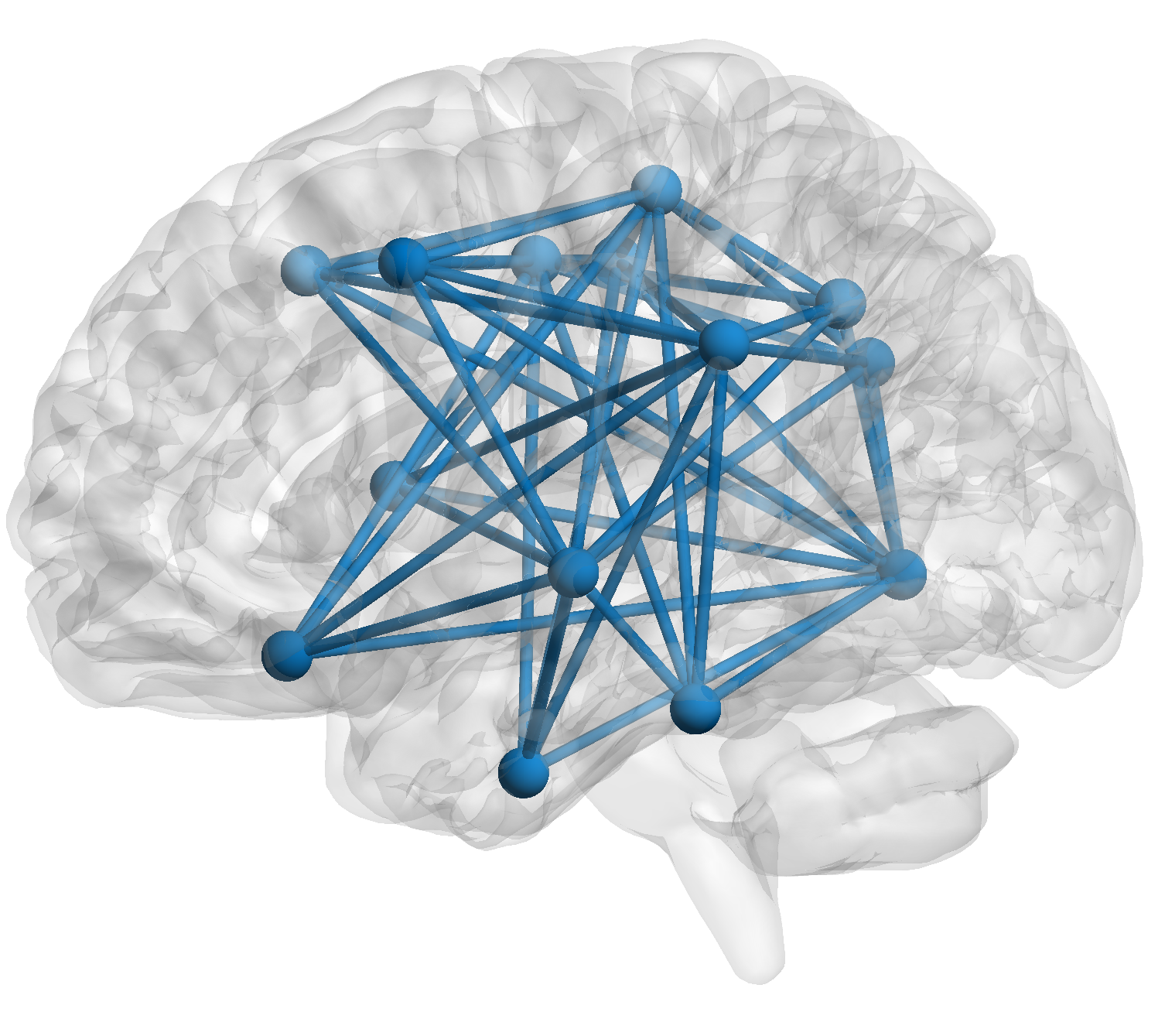}
     \includegraphics[width=.24\textwidth]{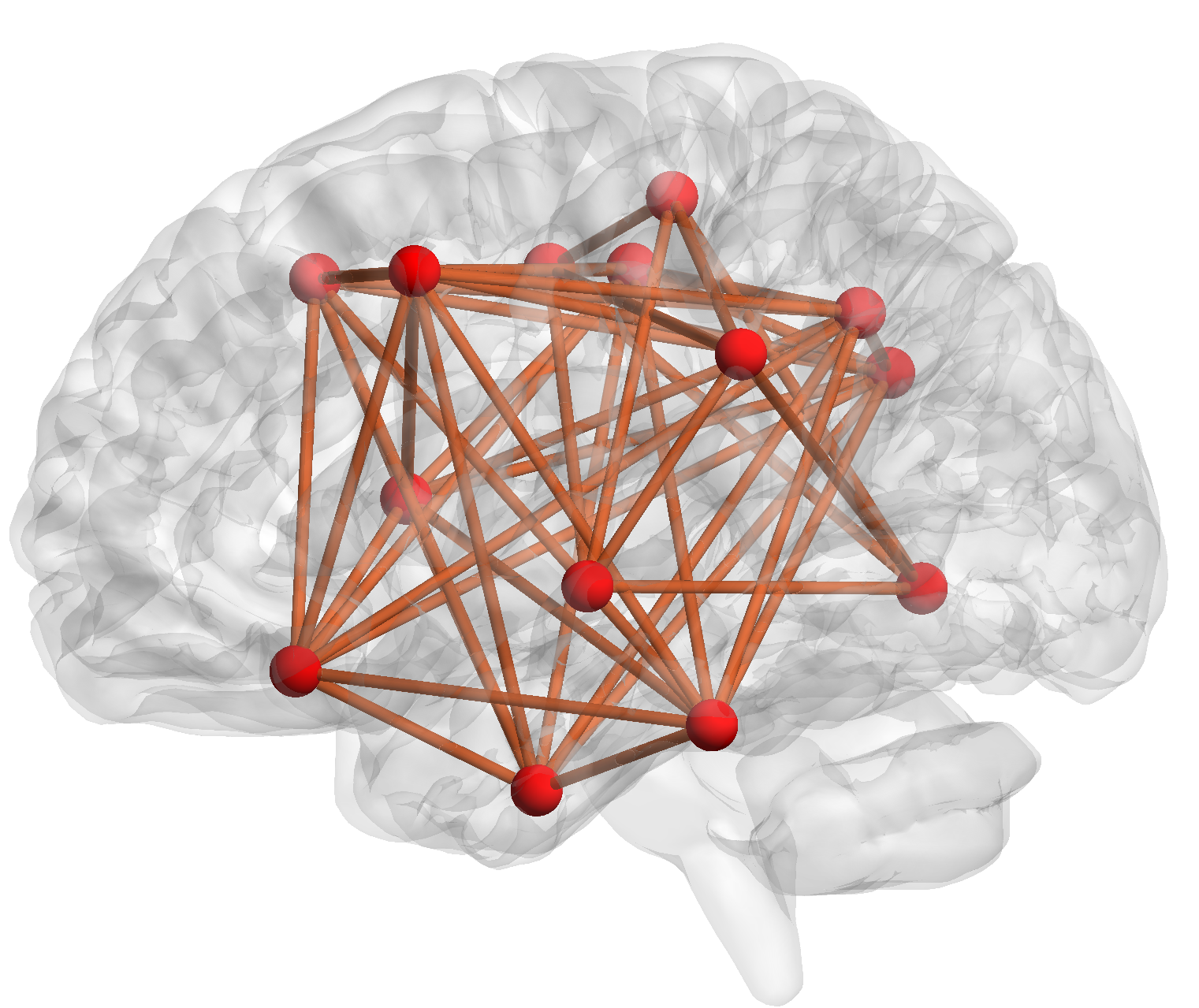}
      \includegraphics[width=.24\textwidth]{figure/legend2.png}
     
    \includegraphics[width=.24\textwidth]{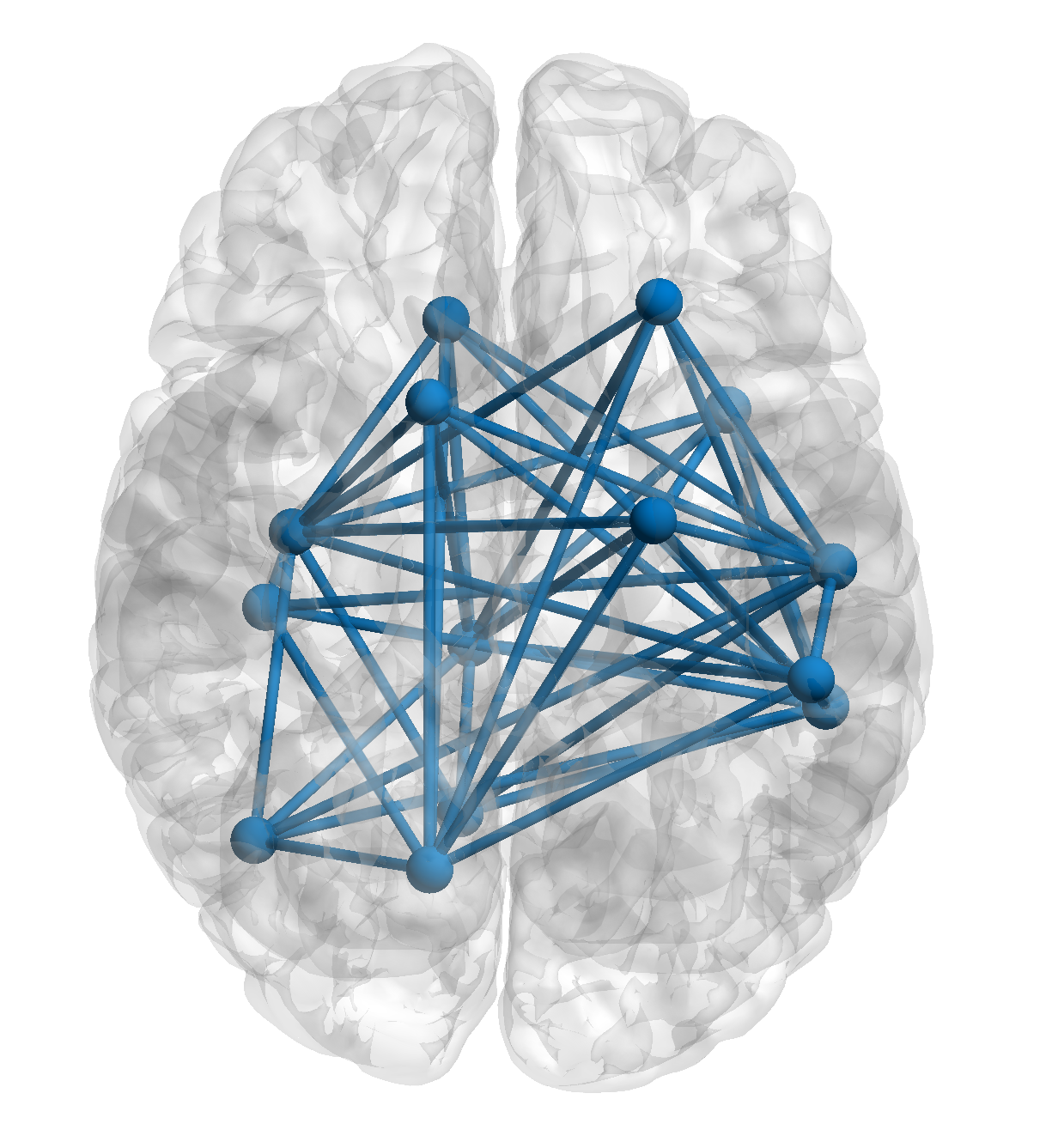}
    \includegraphics[width=.24\textwidth]{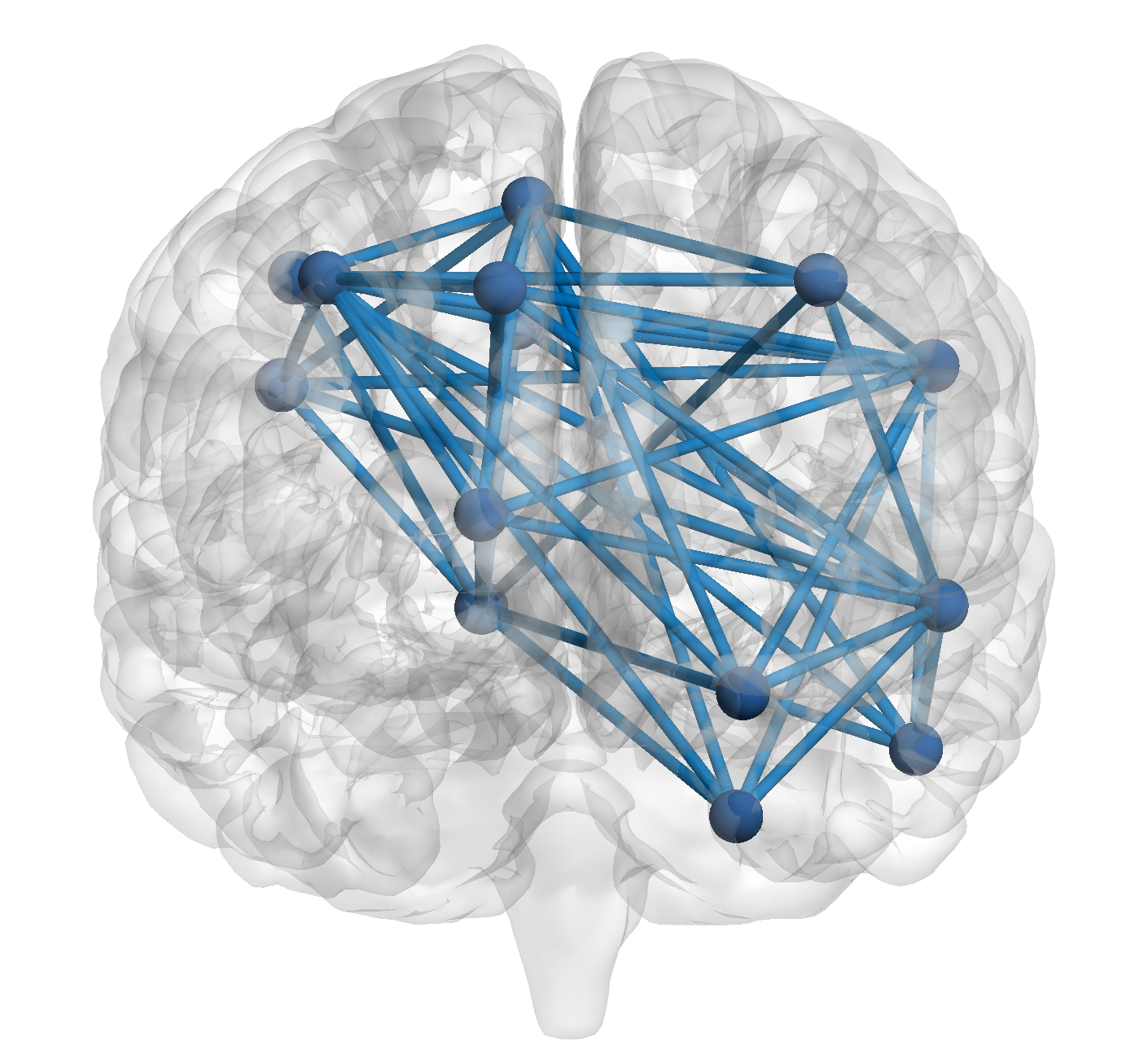}
     \includegraphics[width=.24\textwidth]{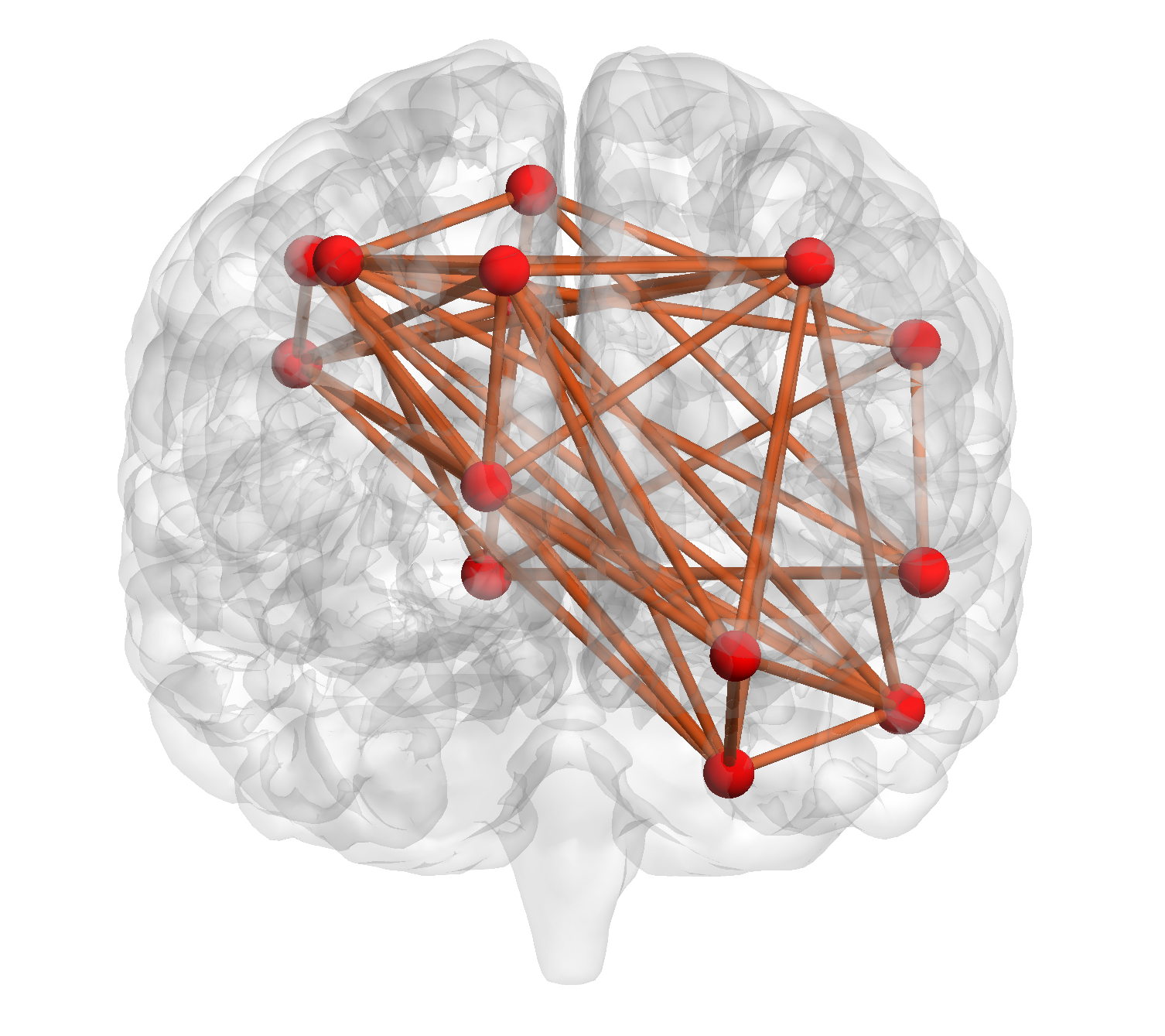}
     \includegraphics[width=.24\textwidth]{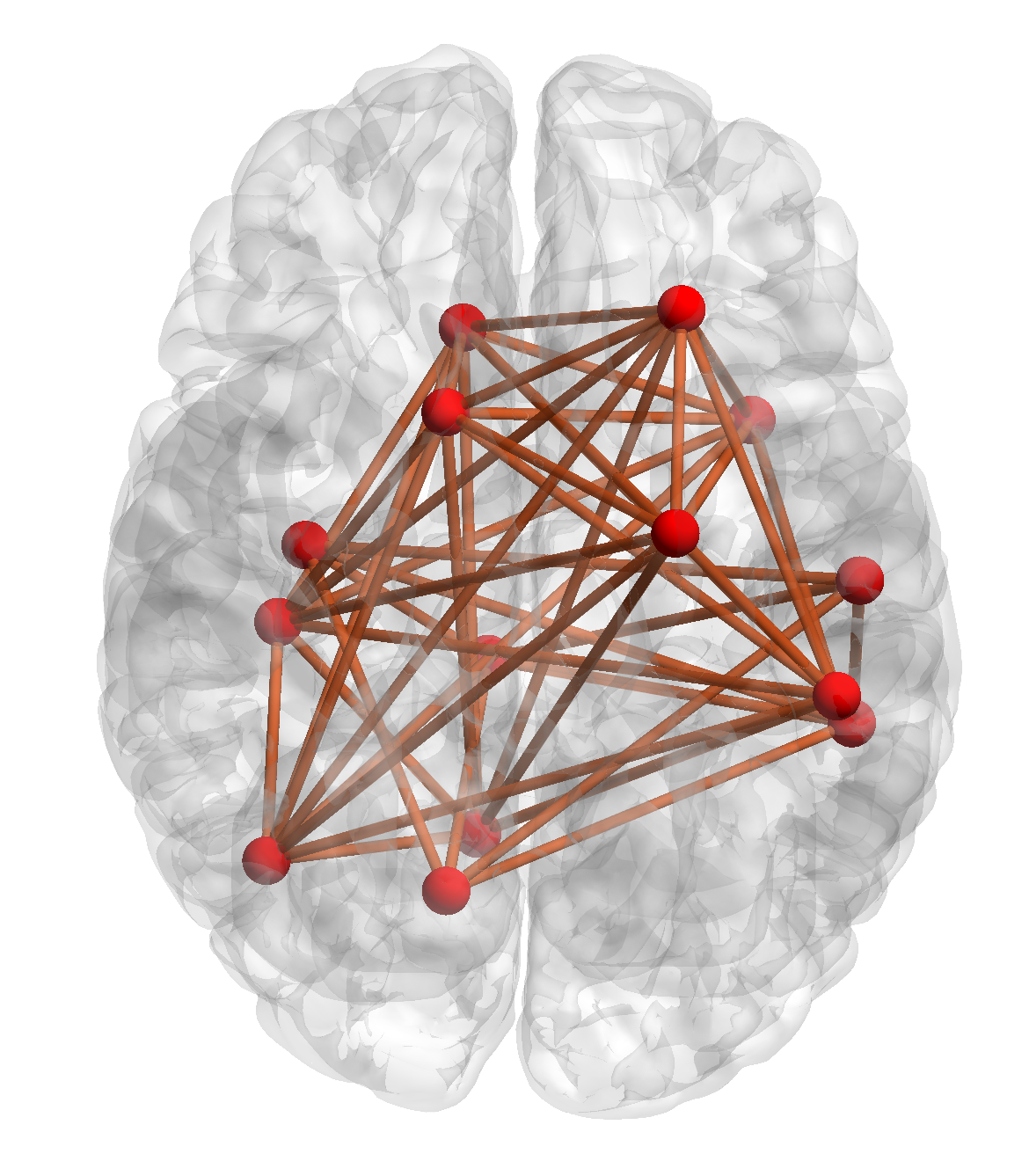}
     
    \caption{The identified sub-networks within the truly mediating subgraphs that are negatively and positively linked with AD survival. 
}\label{fig:foobar}
\end{figure}

Finally, Figure \ref{fig:foobar} shows the components within the truly mediating subgraphs that are negatively and positively linked with the outcome. These components can be summarized into the corresponding positive and negative sub-networks within each subgraph. While those  sub-networks within individual subgraphs are not overlapping, there are common connections within and between canonical functional systems in both positive and negative sub-networks. Of note, such an anti-correlated brain sub-network containing both positive and negative associations with  behavior is widely seen in functional connectivity \citep{yip2019connectome}. Our result indicates it also exists in structural connectivity to carry mediating effects. Overall, both the characterization of this AD survival-related imaging genetics effect mechanism and the cartography of mediating brain network configurations provide a great potential to understand the etiology of the disease and direct future neuronal targets for genetic interventions.

\section{Discussion}\label{sec:dis}
In this paper, motivated by exploring the neurobiological etiology and molecular support for AD survival, we propose a general Bayesian mediation analysis with a time-to-event outcome and a network-variate mediator to dissect the effect pathways among brain structural connectivity, genotype of interest and time to AD onset.
To accommodate the symmetric and hollow architecture of structural connectivity, we develop a symmetric matrix-variate AFT model to characterize the effect pathway from brain connectivity to the outcome, and a symmetric matrix response regression to capture the impact from the genetic exposure to brain connectivity. 
By further assuming structural connectivity operates along different effect paths through separate sets of clique subgraphs, we  impose both within-graph sparsity and between-graph shrinkage to identify informative subgraph configurations and the truly mediating network  elements. Through simulations, we demonstrate the superiority of our method in characterizing different aspects of the effect mechanism. By applying our method to the landmark ADNI study, we uncover the effect mechanism of interest and identify subgraphs within them. These results may provide guidance on understanding disease etiology and prioritizing future intervention targets.

In our modeling framework, we assume that structural connectivity mediating through distinct subgraphs of clique signals when linked with either the genotype or the outcome, and the truly mediating configurations are the network components functioning along both pathways. In the existing literature, there are some alternative options to model the low-rank structure of brain connectivity including  stochastic block modeling \citep{zhao2022bayesian} and non-negative matrix factorization \citep{matrix2018}. Furthermore, it would be useful to extend our work to other settings, as the current modeling framework is able to be embedded within other network-variate parametric assumptions in light of the specific application. Meanwhile, it would be equally interesting to explore more robust nonparametric options \citep{williamson2016nonparametric} to characterize the topological structure.

Another potential future extension of our model is to consider an alternative assumption on the residual term for our AFT model. Currently, we assume it is Normally distributed, but it is likely to be more robust if  the residual term is modeled via a Dirichlet process (DP) mixture model \citep{DPMM}. However, a practical issue of such a modeling choice comes from a growing number of unknown parameters since we need an additional set of parameters to construct DP priors. Given the complex data structure and relatively high-dimensional feature space, increasing modeling complexity may expand the risk of suffering intensive computation and poor posterior mixing with a finite sample size; however, it still merits additional research.

\bibliographystyle{biorefs}
\bibliography{refs}

\begin{thebibliography}{}

\bibitem[\protect\citeauthoryear{Ankit, Ann, Richard, and Danielle}{Ankit
  et~al.}{2018}]{matrix2018}
Ankit, N.~K., Ann, E.~S., Richard, F.~B., and Danielle, S.~B. (2018).
\newblock Modeling and interpreting mesoscale network dynamics.
\newblock {\em Neuroimage} {\bf 180,} 337--349.

\bibitem[\protect\citeauthoryear{Ballester-Plan{\'e}, Schmidt, Laporta-Hoyos,
  Junqu{\'e}, V{\'a}zquez, Delgado, Zubiaurre-Elorza, Macaya, P{\'o}o, Toro,
  et~al\mbox{.}}{Ballester-Plan{\'e} et~al.}{2017}]{ballester2017whole}
Ballester-Plan{\'e}, J., Schmidt, R., Laporta-Hoyos, O., Junqu{\'e}, C.,
  V{\'a}zquez, {\'E}., Delgado, I., Zubiaurre-Elorza, L., Macaya, A., P{\'o}o,
  P., Toro, E., et~al. (2017).
\newblock Whole-brain structural connectivity in dyskinetic cerebral palsy and
  its association with motor and cognitive function.
\newblock {\em Human Brain Mapping} {\bf 38,} 4594--4612.

\bibitem[\protect\citeauthoryear{Barbieri and Berger}{Barbieri and
  Berger}{2004}]{barbieri2004optimal}
Barbieri, M.~M. and Berger, J.~O. (2004).
\newblock Optimal predictive model selection.
\newblock {\em The annals of statistics} {\bf 32,} 870--897.

\bibitem[\protect\citeauthoryear{Bassett and Sporns}{Bassett and
  Sporns}{2017}]{bassett2017network}
Bassett, D.~S. and Sporns, O. (2017).
\newblock Network neuroscience.
\newblock {\em Nature neuroscience} {\bf 20,} 353--364.

\bibitem[\protect\citeauthoryear{Bertram and Tanzi}{Bertram and
  Tanzi}{2008}]{bertram2008thirty}
Bertram, L. and Tanzi, R.~E. (2008).
\newblock Thirty years of alzheimer's disease genetics: the implications of
  systematic meta-analyses.
\newblock {\em Nature Reviews Neuroscience} {\bf 9,} 768--778.

\bibitem[\protect\citeauthoryear{Ch{\'e}n, Crainiceanu, Ogburn, Caffo, Wager,
  and Lindquist}{Ch{\'e}n et~al.}{2018}]{chen2018high}
Ch{\'e}n, O.~Y., Crainiceanu, C., Ogburn, E.~L., Caffo, B.~S., Wager, T.~D.,
  and Lindquist, M.~A. (2018).
\newblock High-dimensional multivariate mediation with application to
  neuroimaging data.
\newblock {\em Biostatistics} {\bf 19,} 121--136.

\bibitem[\protect\citeauthoryear{Chen, Mandal, Zhu, and Liu}{Chen
  et~al.}{2022}]{chen2022imaging}
Chen, T., Mandal, A., Zhu, H., and Liu, R. (2022).
\newblock Imaging genetic based mediation analysis for human cognition.
\newblock {\em Frontiers in neuroscience} {\bf 16,}.

\bibitem[\protect\citeauthoryear{Derkach, Pfeiffer, Chen, and Sampson}{Derkach
  et~al.}{2019}]{derkach2019high}
Derkach, A., Pfeiffer, R.~M., Chen, T.-H., and Sampson, J.~N. (2019).
\newblock High dimensional mediation analysis with latent variables.
\newblock {\em Biometrics} {\bf 75,} 745--756.

\bibitem[\protect\citeauthoryear{Gelman, Rubin, et~al\mbox{.}}{Gelman
  et~al.}{1992}]{gelman1992inference}
Gelman, A., Rubin, D.~B., et~al. (1992).
\newblock Inference from iterative simulation using multiple sequences.
\newblock {\em Statistical science} {\bf 7,} 457--472.

\bibitem[\protect\citeauthoryear{Greene, Gao, Scheinost, and Constable}{Greene
  et~al.}{2018}]{greene2018task}
Greene, A.~S., Gao, S., Scheinost, D., and Constable, R.~T. (2018).
\newblock Task-induced brain state manipulation improves prediction of
  individual traits.
\newblock {\em Nature communications} {\bf 9,} 1--13.

\bibitem[\protect\citeauthoryear{Hashimoto, Ohi, Yamamori, Yasuda, Fujimoto,
  Umeda-Yano, Watanabe, Fukunaga, and Takeda}{Hashimoto
  et~al.}{2015}]{hashimoto2015imaging}
Hashimoto, R., Ohi, K., Yamamori, H., Yasuda, Y., Fujimoto, M., Umeda-Yano, S.,
  Watanabe, Y., Fukunaga, M., and Takeda, M. (2015).
\newblock Imaging genetics and psychiatric disorders.
\newblock {\em Current molecular medicine} {\bf 15,} 168--175.

\bibitem[\protect\citeauthoryear{Hopper and Vogel}{Hopper and
  Vogel}{1976}]{limbic1976}
Hopper, M. and Vogel, F. (1976).
\newblock The limbic system in alzheimer's disease. a neuropathologic
  investigation.
\newblock {\em Am J Pathol} {\bf 85,} 1--20.

\bibitem[\protect\citeauthoryear{Huang and Pan}{Huang and
  Pan}{2016}]{huang_and_pan_2016}
Huang, Y.~T. and Pan, W.~C. (2016).
\newblock Hypothesis test of mediation effect in causal mediation model with
  high-dimensional continuous mediators.
\newblock {\em Biometrics} {\bf 72,} 402--413.

\bibitem[\protect\citeauthoryear{Huang and Yang}{Huang and
  Yang}{2017}]{SurMed2017}
Huang, Y.~T. and Yang, H.~I. (2017).
\newblock Causal mediation analysis of survival outcome with multiple
  mediators.
\newblock {\em Epidemiology (Cambridge, Mass.)} {\bf 28,} 370--378.

\bibitem[\protect\citeauthoryear{Imai, Keele, and Tingley}{Imai
  et~al.}{2010}]{imai2010general}
Imai, K., Keele, L., and Tingley, D. (2010).
\newblock A general approach to causal mediation analysis.
\newblock {\em Psychological methods} {\bf 15,} 309.

\bibitem[\protect\citeauthoryear{Imai and Yamamoto}{Imai and
  Yamamoto}{2013}]{imai2013identification}
Imai, K. and Yamamoto, T. (2013).
\newblock Identification and sensitivity analysis for multiple causal
  mechanisms: Revisiting evidence from framing experiments.
\newblock {\em Political Analysis} {\bf 21,} 141--171.

\bibitem[\protect\citeauthoryear{Jirsa and McIntosh}{Jirsa and
  McIntosh}{2007}]{jirsa2007handbook}
Jirsa, V.~K. and McIntosh, A.~R. (2007).
\newblock {\em Handbook of brain connectivity}, volume~1.
\newblock Springer.

\bibitem[\protect\citeauthoryear{Kenny}{Kenny}{2008}]{mediation2008}
Kenny, D. (2008).
\newblock Reflections on mediation.
\newblock {\em Organizational Research Methods} {\bf 11,} 353--358.

\bibitem[\protect\citeauthoryear{Kuo and Mallick}{Kuo and Mallick}{1997}]{DPMM}
Kuo, L. and Mallick, B. (1997).
\newblock Bayesian semiparametric inference for the accelerated failure‐time
  model.
\newblock {\em Canadian Journal of Statistics} {\bf 25,} 457--472.

\bibitem[\protect\citeauthoryear{Lee, Yoo, Lee, Chung, Lim, Yoon, and
  Jeong}{Lee et~al.}{2016}]{default2016}
Lee, E., Yoo, K., Lee, Y., Chung, J., Lim, J., Yoon, B., and Jeong, Y. (2016).
\newblock Default mode network functional connectivity in early and late mild
  cognitive impairment: Results from the alzheimer's disease neuroimaging
  initiative.
\newblock {\em Alzheimer Dis Assoc Disord} {\bf 30,} 289--296.

\bibitem[\protect\citeauthoryear{Li, Fang, and Contreras}{Li
  et~al.}{2017}]{DTI2017}
Li, H., Fang, S., and Contreras, J. (2017).
\newblock Brain explorer for connectomic analysis.
\newblock {\em Brain Inform} {\bf 4,} 253--269.

\bibitem[\protect\citeauthoryear{Lindquist}{Lindquist}{2012}]{lindquist2012functional}
Lindquist, M.~A. (2012).
\newblock Functional causal mediation analysis with an application to brain
  connectivity.
\newblock {\em Journal of the American Statistical Association} {\bf 107,}
  1297--1309.

\bibitem[\protect\citeauthoryear{Park and Casella}{Park and
  Casella}{2008}]{park2008bayesian}
Park, T. and Casella, G. (2008).
\newblock The bayesian lasso.
\newblock {\em Journal of the American Statistical Association} {\bf 103,}
  681--686.

\bibitem[\protect\citeauthoryear{Power, Cohen, Nelson, Wig, Barnes, Church,
  Vogel, Laumann, Miezin, Schlaggar, et~al\mbox{.}}{Power
  et~al.}{2011}]{power2011functional}
Power, J.~D., Cohen, A.~L., Nelson, S.~M., Wig, G.~S., Barnes, K.~A., Church,
  J.~A., Vogel, A.~C., Laumann, T.~O., Miezin, F.~M., Schlaggar, B.~L., et~al.
  (2011).
\newblock Functional network organization of the human brain.
\newblock {\em Neuron} {\bf 72,} 665--678.

\bibitem[\protect\citeauthoryear{Schwarz, Gozzi, and Bifone}{Schwarz
  et~al.}{2008}]{schwarz2008community}
Schwarz, A.~J., Gozzi, A., and Bifone, A. (2008).
\newblock Community structure and modularity in networks of correlated brain
  activity.
\newblock {\em Magnetic resonance imaging} {\bf 26,} 914--920.

\bibitem[\protect\citeauthoryear{Shen, Finn, Scheinost, Rosenberg, Chun,
  Papademetris, and Constable}{Shen et~al.}{2017}]{shen2017using}
Shen, X., Finn, E.~S., Scheinost, D., Rosenberg, M.~D., Chun, M.~M.,
  Papademetris, X., and Constable, R.~T. (2017).
\newblock Using connectome-based predictive modeling to predict individual
  behavior from brain connectivity.
\newblock {\em nature protocols} {\bf 12,} 506--518.

\bibitem[\protect\citeauthoryear{Simpson, Hayasaka, and Laurienti}{Simpson
  et~al.}{2011}]{simpson2011exponential}
Simpson, S.~L., Hayasaka, S., and Laurienti, P.~J. (2011).
\newblock Exponential random graph modeling for complex brain networks.
\newblock {\em PloS one} {\bf 6,} e20039.

\bibitem[\protect\citeauthoryear{Song, Zhou, Zhang, Zhao, Liu, Kardia, Roux,
  Needham, Smith, and Mukherjee}{Song et~al.}{2020}]{song2020bayesian}
Song, Y., Zhou, X., Zhang, M., Zhao, W., Liu, Y., Kardia, S.~L., Roux, A.
  V.~D., Needham, B.~L., Smith, J.~A., and Mukherjee, B. (2020).
\newblock Bayesian shrinkage estimation of high dimensional causal mediation
  effects in omics studies.
\newblock {\em Biometrics} {\bf 76,} 700--710.

\bibitem[\protect\citeauthoryear{Sun, Chang, Zhao, and Long}{Sun
  et~al.}{2018}]{sun2018knowledge}
Sun, W., Chang, C., Zhao, Y., and Long, Q. (2018).
\newblock Knowledge-guided bayesian support vector machine for high-dimensional
  data with application to analysis of genomics data.
\newblock In {\em 2018 IEEE International Conference on Big Data (Big Data)},
  pages 1484--1493. IEEE.

\bibitem[\protect\citeauthoryear{Tang, Holland, Dale, and Miller}{Tang
  et~al.}{2015}]{apoe2015}
Tang, X., Holland, D., Dale, A., and Miller, M. (2015).
\newblock Alzheimer's disease neuroimaging initiative. apoe affects the volume
  and shape of the amygdala and the hippocampus in mild cognitive impairment
  and alzheimer's disease.
\newblock {\em J Alzheimers Dis} {\bf 47,} 645--660.

\bibitem[\protect\citeauthoryear{VanderWeele}{VanderWeele}{2011}]{vanderweele2011causal}
VanderWeele, T.~J. (2011).
\newblock Causal mediation analysis with survival data.
\newblock {\em Epidemiology (Cambridge, Mass.)} {\bf 22,} 582.

\bibitem[\protect\citeauthoryear{Wang, Lin, Cole, and Zhang}{Wang
  et~al.}{2021}]{wang2021learning}
Wang, L., Lin, F.~V., Cole, M., and Zhang, Z. (2021).
\newblock Learning clique subgraphs in structural brain network classification
  with application to crystallized cognition.
\newblock {\em Neuroimage} {\bf 225,} 117493.

\bibitem[\protect\citeauthoryear{Wang, Nelson, and Albert}{Wang
  et~al.}{2013}]{wang_2013}
Wang, W., Nelson, S., and Albert, J.~M. (2013).
\newblock Estimation of causal mediation effects for a dichotomous outcome in
  multiple-mediator models using the mediation formula.
\newblock {\em Statistics in medicine} {\bf 32,} 4211--4228.

\bibitem[\protect\citeauthoryear{Weiler, Campos, Nogueira, Damasceno, Cendes,
  and Balthazar}{Weiler et~al.}{2014}]{withinconnect}
Weiler, M., Campos, B., Nogueira, M., Damasceno, B., Cendes, F., and Balthazar,
  M. (2014).
\newblock Structural connectivity of the default mode network and cognition in
  alzheimer's disease.
\newblock {\em Psychiatry Research: Neuroimaging} {\bf 223,} 15--22.

\bibitem[\protect\citeauthoryear{Weiner, Veitch, Aisen, Beckett, Cairns, Green,
  Harvey, Jack, Jagust, Liu, Morris, Petersen, Saykin, Schmidt, Shaw, Siuciak,
  Soares, Toga, and Trojanowski}{Weiner et~al.}{2011}]{ADNI2011}
Weiner, M.~W., Veitch, D.~P., Aisen, P.~S., Beckett, L.~A., Cairns, N.~J.,
  Green, R.~C., Harvey, D., Jack, C.~R., Jagust, W., Liu, E., Morris, J.~C.,
  Petersen, R.~C., Saykin, A.~J., Schmidt, M.~E., Shaw, L., Siuciak, J.~A.,
  Soares, H., Toga, A.~W., and Trojanowski, J.~Q. (2011).
\newblock The alzheimer's disease neuroimaging initiative: a review of papers
  published since its inception.
\newblock {\em Alzheimer's \& dementia : the journal of the Alzheimer's
  Association} {\bf 8,} S1--S68.

\bibitem[\protect\citeauthoryear{Williamson}{Williamson}{2016}]{williamson2016nonparametric}
Williamson, S.~A. (2016).
\newblock Nonparametric network models for link prediction.
\newblock {\em The Journal of Machine Learning Research} {\bf 17,} 7102--7121.

\bibitem[\protect\citeauthoryear{Yip, Scheinost, Potenza, and Carroll}{Yip
  et~al.}{2019}]{yip2019connectome}
Yip, S.~W., Scheinost, D., Potenza, M.~N., and Carroll, K.~M. (2019).
\newblock Connectome-based prediction of cocaine abstinence.
\newblock {\em American Journal of Psychiatry} {\bf 176,} 156--164.

\bibitem[\protect\citeauthoryear{Zeng, Shao, and Zhou}{Zeng
  et~al.}{2021}]{zeng2021statistical}
Zeng, P., Shao, Z., and Zhou, X. (2021).
\newblock Statistical methods for mediation analysis in the era of
  high-throughput genomics: current successes and future challenges.
\newblock {\em Computational and structural biotechnology journal} {\bf 19,}
  3209--3224.

\bibitem[\protect\citeauthoryear{Zhao, Chen, Cai, Lichenstein, Potenza, and
  Yip}{Zhao et~al.}{2022}]{zhao2022bayesian}
Zhao, Y., Chen, T., Cai, J., Lichenstein, S., Potenza, M., and Yip, S. (2022).
\newblock Bayesian network mediation analysis with application to brain
  functional connectome.
\newblock {\em Statistics in Medicine} {\bf 41,} 3991--4005.

\bibitem[\protect\citeauthoryear{Zhao, Li, and Caffo}{Zhao
  et~al.}{2021}]{zhao2021multimodal}
Zhao, Y., Li, L., and Caffo, B.~S. (2021).
\newblock Multimodal neuroimaging data integration and pathway analysis.
\newblock {\em Biometrics} {\bf 77,} 879--889.

\bibitem[\protect\citeauthoryear{Zhao and Luo}{Zhao and
  Luo}{2019}]{zhao2019granger}
Zhao, Y. and Luo, X. (2019).
\newblock Granger mediation analysis of multiple time series with an
  application to functional magnetic resonance imaging.
\newblock {\em Biometrics} {\bf 75,} 788--798.

\end{thebibliography}

\end{document}